\documentclass[fleqn,usenatbib]{mnras}

\usepackage{newtxtext,newtxmath}
\usepackage{physics}

\usepackage[T1]{fontenc}

\usepackage{mhchem}

\DeclareRobustCommand{\VAN}[3]{#2}
\let\VANthebibliography\thebibliography
\def\thebibliography{\DeclareRobustCommand{\VAN}[3]{##3}\VANthebibliography}

\usepackage{graphicx}	
\usepackage{amsmath}	
\usepackage{orcidlink}

\newcommand{\PUTh}{P^\mathrm{U/Th}}
\newcommand{\pWkgMg}{\mathrm{pW~(kg~Mg)}^{-1}}
\newcommand{\perplex}{\textsc{Perple\_X}}

\defcitealias{basant25}{B25}
\defcitealias{jahandar24}{J24}

\usepackage{xcolor}

\title[
    The Barnard's Star Planetary System
]{
    The Barnard's Star Planetary System: Stability, Composition, and Evolution of Four Sub-Earth Exoplanets
}

\author[X. Byrne et al.]{
    Xander Byrne
    \orcidlink{0000-0001-9488-238X},$^{1}$\thanks{E-mail: \href{mailto:xbyrne@ast.cam.ac.uk}{xbyrne@ast.cam.ac.uk}}
    Claire Marie Guimond
    \orcidlink{0000-0003-1521-5461},$^{2,3}$
    Amy Bonsor
    \orcidlink{0000-0002-8070-1901},$^{1}$
    Haiyang S.\ Wang
    \orcidlink{0000-0001-8618-3343},$^{4}$
    \newauthor
    \ Sophia R.\ Vaughan
    \orcidlink{0000-0002-8199-9818},$^{5}$
    James G. Rogers
    \orcidlink{0000-0001-7615-6798}$^{1}$
    \\
$^{1}$Institute of Astronomy, University of Cambridge, Madingley Road, Cambridge CB3 0HA, UK\\
$^{2}$Atmospheric, Oceanic, and Planetary Physics, Department of Physics, University of Oxford, Sherrington Rd, Oxford, OX1 3PU, UK\\
$^{3}$Department of Earth and Planetary Sciences, ETH Zurich, Sonneggstrasse 5, 8092 Zurich, Switzerland\\
$^{4}$Centre for Star and Planet Formation, Globe Institute, University of Copenhagen, Øster Voldgade 5-7, 1350 Copenhagen, Denmark\\
$^{5}$Max-Planck-Institut für Astronomie, Königstuhl 17, 69117 Heidelberg, Germany
}

\date{Accepted XXX. Received YYY; in original form ZZZ}

\pubyear{\the\year{}}

\begin{document}
\label{firstpage}
\pagerange{\pageref{firstpage}--\pageref{lastpage}}
\maketitle

\begin{abstract}
Barnard's Star is the nearest single star to the Sun ($1.8\,\text{pc}$), and hosts four recently-discovered planets.
The star also has well-characterized stellar abundances of important rock-forming elements, including Fe, Mg, and Si.
For refractory elements like these, the planets have likely inherited similar bulk elemental abundance ratios to the star, facilitating modelling of their interior structures.
We present here an analysis of the Barnard's Star planetary system on several fronts.
We perform a detailed stability analysis of the system, ascertaining that all four planets likely have masses between $0.19$ and $0.84~M_{\earth}$, and are likely tidally locked, whereas a 4:3 mean-motion resonance chain for the inner three planets cannot be ruled out.
Using atmospheric evolution models, we show that the prospect of extant primary atmospheres is highly unlikely on any of the planets.
Barnard's Star's abnormally high Mg/Si ratio and low Th/Mg ratio imply planetary mantles which (a) are rich in (Mg,Fe)O ferropericlase; (b) have less than half the water capacity as Earth; (c) generate about half of the radiogenic heating as Earth; and (d) are cool and unlikely to have outgassed secondary atmospheres.
Our analysis of this system presents an accessible set of first steps for the study of other nearby exoplanetary systems, as well as sub-Earth planets which will be increasingly discovered over the coming years.
\end{abstract}




\section{Introduction}
\label{sec:intro}

M dwarfs are the dominant stellar class in our Galaxy \citep[e.g.][]{henry94, chabrier00}.
Notwithstanding biases of exoplanet detection methods towards smaller host stars, M dwarf planetary systems are ubiquitous \citep[e.g.][]{kopparapu13occ, dressing15}.

Barnard's Star is an M4 dwarf at a distance of $1.828~\text{pc}$, the nearest single star to the Sun (further stellar parameters are presented in Table~\ref{tab:star}).
Following a controversial history of exoplanet detection claims \citep{vandekamp63, gatewood73, gatewood95, ribas18, lubin21}, a compact system of four planets has recently been confirmed using the radial velocity (RV) technique (\citealt{gonzalezhernandez24, basant25}, hereafter \citetalias{basant25}).
The planets each have sub-Earth minimum masses and orbital periods on the order of days (Table~\ref{tab:planets}).

\begin{table}
\caption{
    Selected parameters of Barnard's Star.
}
\label{tab:star}
\begin{tabular}{llllc}
\hline
Parameter & & Value & Error & Ref. \\
\hline
Distance & [pc] & $1.82823$ & $0.00013$ & (1) \\
Proper motion & [mas~yr$^{-1}$] & $10~393.35^*$ & $0.04$ & (1) \\
\textit{G} & [mag] & 8.194 & 0.003 & (1) \\
Mass & [$M_{\sun}$] & $0.159$ & 0.016 & (2) \\
Radius & [$R_{\sun}$] & $0.1869$ & $0.0012$ & (2) \\
Spectral Class & & M4.2 && (2) \\
$T_\mathrm{eff}$ & [K] & $3238$ & 11 & (2) \\
Rotation & [d] & $145$ & $15$ & (3) \\
$\mathrm{[Fe/H]}$ & & $-0.38$ & $0.03$ & (4) \\
Age & [Gyr] & 10 & 3 & (5) \\
\hline
\end{tabular}\\
{\raggedright
    $^*$Fastest of any known star.
    (1) \citet{gaiadr3}; (2) \citet{mann13}; (3) \citet{toledopatron19}; (4) \citet{jahandar24}; (5) \citet{gauza15}
}
\end{table}

\begin{table*}
\caption{
    Selected parameters of the Barnard's Star planets, as fitted from RV observations \citepalias{basant25}.
    The slightly awkward naming convention of the planets is in order of RV amplitude $K$ as estimated in \citet{gonzalezhernandez24}, though note that these values have since been refined and the names are no longer in order of this quantity.
    They are listed here in order of orbital period.
}
\label{tab:planets}
\begin{tabular}{lllllllll}
\hline
Parameter & d && b && c && e\\
\hline
$a$ [au] & $0.0188$ & $\pm0.0003$ & $0.0229$ & $\pm0.0003$ & $0.0274$ & $\pm0.0004$ & $0.0382$ & $\pm0.0005$ \\
$P$ [days] & $2.3402$ & $\pm0.0003$ & $3.1542$ & $\pm0.0004$ & $4.1244$ & $\pm0.0006$ & $6.7392$ & $\pm0.0028$ \\
$e$ & $0.04$ & $^{+0.05}_{-0.03}$ & $0.03$ & $^{+0.03}_{-0.02}$ & $0.08$ & $^{+0.06}_{-0.05}$ & $0.04$ & $^{+0.04}_{-0.03}$ \\
$M \sin i$ [$M_{\earth}]$ & $0.263$ & $\pm0.024$ & $0.299$ & $\pm0.026$ & $0.335$ & $\pm0.030$ & $0.193$ & $\pm0.033$ \\
$K$ [m~s$^{-1}$] & $0.428$ & $\pm0.036$ & $0.440$ & $\pm0.036$ & $0.452$ & $\pm0.038$ & $0.221^*$ & $\pm0.037$ \\
\hline
\end{tabular}\\
{\raggedright
    $^*$Lowest of any confirmed RV-discovered planet to date.
}
\end{table*}

Barnard's Star's `habitable zone' \citep{kopparapu13hab, kopparapu14} corresponds to periods between 10 and 42~days; its `abiogenesis zone' -- the distance from a star within which the ultraviolet (UV) flux is sufficient to photochemically initiate prebiotic chemical reactions \citep{rimmer18} -- is within 1.2~days.
The planets are therefore expected to be too hot to support life, as well as too distant to initiate it anyway.
While life as we know it is therefore unlikely to develop on these planets, they remain of great interest due to their proximity: these are currently the 3rd, 4th, 5th and 6th nearest confirmed exoplanets.

Planets ultimately form from the same molecular cloud as their host star, and are thus compositionally related.
Refractory elements (with condensation temperatures $T_C\gtrsim1200~\text{K}$) solidify early in the evolution of protoplanetary discs, and are thus incorporated into planets in approximately stellar ratios in most cases (but see e.g.\ \citealt{zaveri26}).
Some of these ratios (e.g.\ Mg/Si and Fe/Si) control the internal structure, mineralogy, and geodynamics of a planet, and these ratios are expected to faithfully reflect those of their stars within observational uncertainty \citep{thiabaud15, dorn17, hinkel_star_2018, putirka_composition_2019, guimond_stars_2024, sanchez25}.
Indeed the solar ratios of these elements are reflected in the Earth \citep[e.g.,][]{javoy10}, and Mars \citep{yoshizaki_composition_2020}
(though not Mercury, whose super-solar Fe/Si may indicate a unique formation regime \citep{lewis72} or collisional processing; \citealt{benz07}).
Due to the propensity of certain mineral phases to incorporate hydroxyl \citep{smyth87, kohlstedt96}, once a planet's mantle mineralogy has been ascertained from its bulk elemental composition, the water storage capacity of the planet's mantle can also be calculated \citep{dong_water_2022, guimond_mantle_2023, boley_underground_2025}.
Furthermore, the Th/Mg ratio enables estimates of the radiogenic heat generated in the mantles \citep{wang22}.
Thorium is an important long-lived radioactive element whose decay significantly contributes to the Earth's internal heat budget \citep[e.g.,][]{oneill20} and hence its tectonic regime \citep{oneill20gce}, geodynamo and magnetic field \citep{nimmo20}, and ultimately habitability \citep{lammer09}.
These abundance ratios can sometimes be measured from stellar spectra, permitting inferences on the geology, volatile inventory, and thermal evolution of their planets.

A detailed chemical analysis of Barnard's Star has been conducted (\citealt{jahandar24}, hereafter \citetalias{jahandar24}) from high-resolution high-signal-to-noise near-infrared ($R\sim70\,000$; $\mathrm{S/N}\sim1000$; $0.97$--$2.49~\mu\mathrm{m}$) spectra obtained using SPIRou \citep{donati20}.
Although M dwarf spectra are generally more difficult to determine abundances from than FGK stars due to broad molecular features, Barnard's Star's low metallicity (${\rm [Fe/H]}=-0.38$; \citetalias{jahandar24}) reduces the severity of line blending, facilitating these abundance measurements.
Abundances of refractory elements Fe, Mg, Si, Ca, Al, Ti, Cr, Sc, V, Y, and Th are all measured.
Na and K are also measured; although these elements are important for some geological processes such as crust production, their lower condensation temperatures ($T_C\approx1000~\text{K}$) mean they are likely depleted in planets relative to refractory elements.
While the volatile elements C and O are also measured, rocky planet C/O ratios are only weakly linked to those of their stars \citep{thiabaud15}.

The elemental composition, interior structure, and geochemistry of a hypothetical Earth-sized habitable-zone planet around $\alpha$~Centauri~AB is investigated in \citet{wang22}, based on stellar abundances derived from HARPS spectra \citep{morel18, wang20}.
The present work has similar goals: to indirectly characterize the Barnard's Star planets based on stellar abundances.
Whereas no planets of $\alpha$ Centauri AB have been confirmed to date, Barnard's Star hosts four confirmed planets.

Our task is however made more challenging by the fact that the planets are in many ways not Earth-like (e.g.\ orbital separation, host stellar type), meaning some assumptions must be made regarding the conditions of their protoplanetary disc.
Furthermore, the unique composition of Mercury demonstrates that the star-planet connection is not fully reliable, and the wildly different atmospheric regimes of the otherwise-similar Venus and Earth demonstrate that detailed surface conditions are generally beyond deduction merely from the stellar abundances, due primarily to uncertainties in the behaviour of volatile elements.
The planets' radii are also uncertain as they are not transiting, but their low masses and high irradiation strongly suggest that they are small and rocky.

While only minimum masses are measured directly from RV measurements, estimated \textit{upper} limits on the masses are presented in Section~\ref{sec:stability}.
Section~\ref{sec:atm_loss} presents the results of atmospheric evolution modelling of the planets; the interior structures and water storage capacities of their mantles are calculated in Section~\ref{sec:interior}.
The implications of Barnard's Star's Th measurement on the planets' internal radiogenic heating and thermal evolution are outlined in Section~\ref{sec:radiogenics}.
We discuss all the above investigations in Section~\ref{sec:discussion}, and conclude in Section~\ref{sec:conclusion}.

\section{Stability analysis and estimated upper mass limits}
\label{sec:stability}

The Barnard's Star planets were discovered using the RV technique, which does not reveal a planet's mass $M$, but rather $M_\mathrm{min} = M \sin i$, where $i$ is the inclination of the planet's orbit to the plane of the sky.
Thus the RV technique alone can only reveal the \textit{minimum} masses of exoplanets, leaving the true mass degenerate with the orbital inclination.

A number of methods have been employed to resolve this degeneracy for non-transiting planets.
Methods that have been applied in other systems include astrometry \citep[e.g.][]{kiefer19}, rotational broadening \citep[e.g.][]{bowler23} photometric light curve fitting \citep{brown09}, analysis of the stellar activity \citep{dumusque14}, or direct measurement of the planetary RV signature \citep{brogi12, rodler12}.

However, none of these methods is feasible for the Barnard's Star planetary system.
Detecting the astrometric signature of Barnard's Star~b alone has been shown to be unlikely with the \textit{Gaia} and \textit{Hubble} space telescopes, even when a much higher mass of the planet is considered \citep{tal-or19}.
Ultimately the planetary system is too low-mass and compact for current astrometric facilities -- or for that matter the photometric light curve or planetary RV methods -- to break the degeneracy.
Identifying inclination from stellar activity is prohibitively more challenging for M dwarfs than for solar-type stars \citep{dumusque14soap}.
The rotational broadening for Barnard's Star has been constrained to be $v_*\sin i < 2~\mathrm{km}~\mathrm{s}^{-1}$ \citep{reiners18}; using the radius and rotation rate in Table~\ref{tab:star}, this corresponds to $\sin i<31$, which does not constrain $i$.

An upper limit on the masses can however be deduced based on the stability of the system \citepalias{basant25}.
The Barnard's Star system is highly compact, with all four planets within $0.04~\text{au}$ of the star.
Even assuming minimum masses, the separations between the planets in terms of their mutual Hill radii are $(\Delta_{\rm db}, \Delta_{\rm bc}, \Delta_{\rm ce})=(13.0, 11.4, 22.0)$.
These are not much larger than the heuristic stability limit of 10 found in \citet{chambers96}.
Furthermore, for planets labelled p and q, $\Delta_{\rm pq}\propto (M_{\rm p} + M_{\rm q})^{-1/3}$, so if the planets' masses are larger than their minimum masses then the Hill separations would be even smaller.
Assuming that the planets' orbits are coplanar, they share an inclination $i$, and their masses are then given by $M = M_{\mathrm{min}} / \sin i$.
If the inclination of the system is low, the planets would all be very massive and the system would likely be unstable over the star's $\sim10~\text{Gyr}$ age \citep{gauza15}.

The dynamical stability of a planetary system can be quickly evaluated using the machine-learning-based \texttt{SPOCK} package \citep{tamayo20}.
The \texttt{SPOCK} algorithm takes as input the stellar mass, planetary masses and orbital parameters (encapsulated in a \texttt{REBOUND} simulation; \citealt{rebound}).
It returns an estimate of the probability that the system is dynamically stable, for at least $10^9$ orbits of the innermost planet.
By scaling up the masses of the Barnard's Star planets from the minimum masses, \citetalias{basant25} report that for planetary masses larger than three times the minimum masses, the stability probability falls below $0.9$.
\citetalias{basant25} hence find that the system is stable only if the system's inclination $i>\arcsin(1/3)\approx 20\degr$.
Notably, it is also found that the eccentricities which best fit the RV data are not long-term stable, even with masses equal to the minimum masses (i.e.\ $i=90\degr$).

We repeat this analysis in finer detail, calculating the stability probability 
for 90 values of $\sin i$, between 0.1 and 1.0, at intervals of 0.01.
For a given value of $\sin i$ we use \texttt{SPOCK} to estimate the stability probability for the Barnard's Star system with planetary masses equal to $M=M_{\rm min}/\sin i$, where the minimum masses are taken from \citetalias{basant25} (also given in Table~\ref{tab:planets}).
We again consider the case where the planets each have eccentricities, arguments of periastron, and times of periastron which best fit the RV data, though we note that these values are poorly constrained \citepalias{basant25}.
We also analyse the case where each planet has a circular orbit (consistent with the RV data at $1.5\sigma$).
To make the parameter space a reasonable size for this brief analysis, we also assume the system to be flat, i.e.\ with no mutual inclinations\footnote{
    We defer a more in-depth Bayesian exploration of the parameter space of compact RV systems to future work (Byrne \& Bonsor, in preparation).
}.

\subsection{Results}
\label{sec:stability_results}

\begin{figure}
\includegraphics[width=\columnwidth]{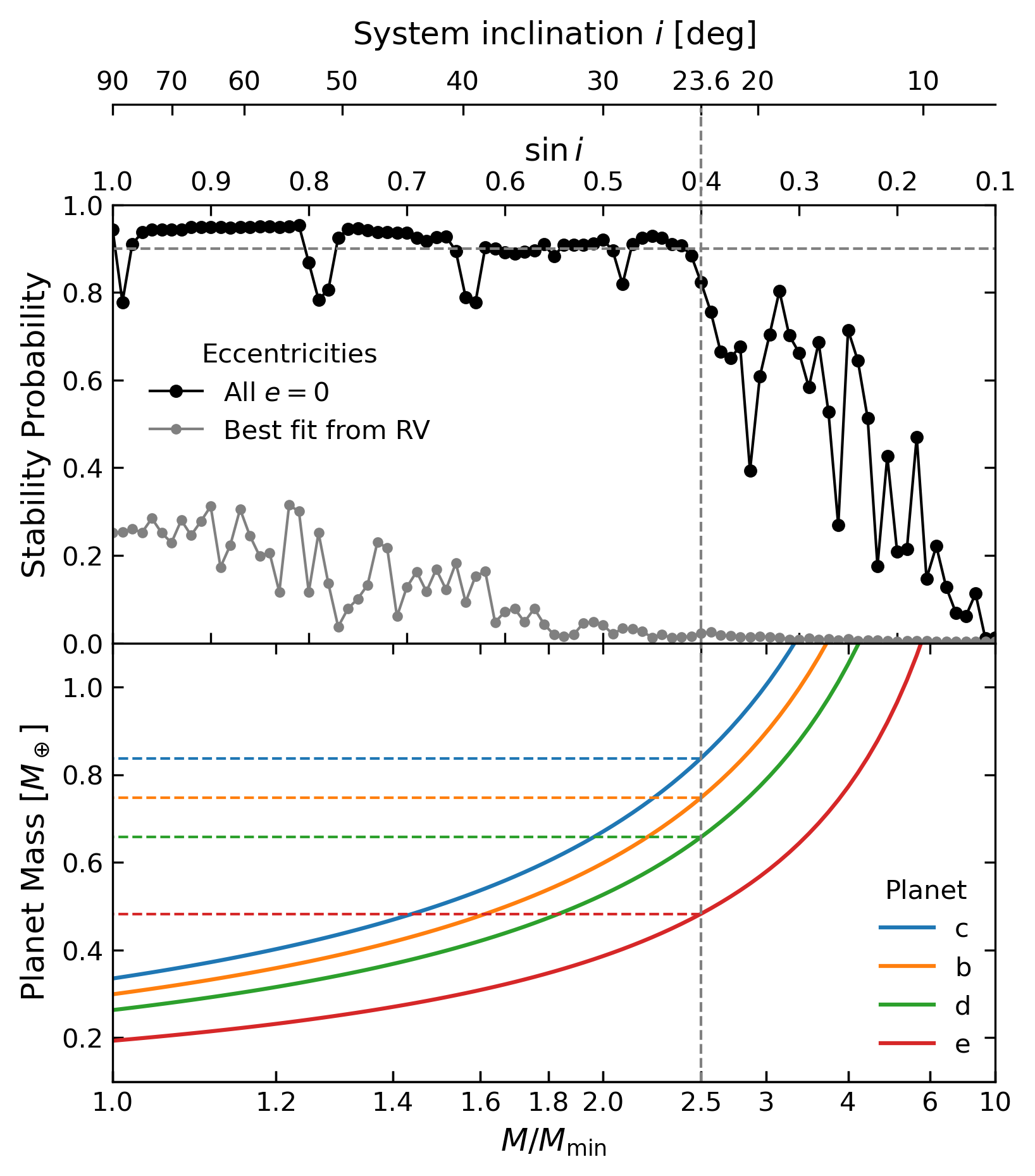}
\caption{
    Stability and masses of the Barnard's Star planetary system as a function of system inclination $i$ (assuming a flat system).
    The upper panel shows the probability that the system is stable over $10^9$ orbits of the innermost planet, d.
    This probability is shown for two cases: (i) the eccentricities are all 0; (ii) the eccentricities are the best-fitting values given in \citetalias{basant25}.
    The system is only stable (with 90\% probability)
    if $i\gtrsim24\degr$ and if the eccentricities are all 0.
    This corresponds to the planets having masses at most 2.5 times their minimum masses, which can be read off the lower panel; the planets are labelled in mass order.
}
\label{fig:stability}
\end{figure}

The results of this analysis are shown in the upper panel of Fig.~\ref{fig:stability}.
If the eccentricities which best fit the RV data are used, we find the system unlikely to be stable for any inclination, corroborating \citetalias{basant25}.
Assuming circular orbits, we find that stable orbits are possible for $\sin i \gtrsim 0.40$, corresponding to $i\gtrsim24\degr$, corresponding to planetary masses up to $2.5\times$ their minimum masses.
(This is a tighter constraint than \citetalias{basant25}, in which a higher factor of 3 is reported in their preliminary analysis.)
This allows us to place soft upper limits on the masses of the planets, as shown in the lower panel of Fig.~\ref{fig:stability}, and reported in Table~\ref{tab:stability}.
One full $N$-body simulation of the system is also conducted, in which the planets have $2.5\times$ their minimum masses; this simulation shows stable behaviour for the timescale explored by \texttt{SPOCK} (Appendix~\ref{app:Nbody}).

Unlike the strict lower bound on the planetary masses represented by the RV minimum mass, the larger masses obtained above at $i=24\degr$ are only soft upper limits, for reasons we outline here.
Firstly, the chaotic nature of many-body gravitational interactions means that only the \textit{probability} of dynamical stability is calculated.
Whereas there may only be a 50\% probability that the system is stable if in fact $i=10\degr$ (Fig.~\ref{fig:stability}), this is hardly impossible.
Secondly, we have explored only a limited parameter space, in which the planets have particular eccentricities, arguments of periastron, and mean anomalies, as well as no mutual inclinations.
It is possible that freeing these parameters could enable stability with higher masses.
Conversely, we note that lower inclination angles are geometrically disfavoured: an isotropic prior on $i$ would be distributed as $\pi(i)=\sin i$.

For these reasons, we interpret these masses -- which we will refer to as `upper masses' -- as probabilistic upper limits on the masses of the Barnard's Star planets.
It should be borne in mind that higher masses are possible, though unlikely.
None the less, this analysis reveals that all four planets are almost certainly rocky, and significantly sub-Earth in mass: the most massive planet in the system (c) has an upper mass of $0.84~M_{\earth}$.

\begin{table}
\caption{
    Mass constraints on the Barnard's Star planets.
    Minimum masses are obtained from the RV measurements in \citetalias{basant25}.
    Soft upper limits on the masses ($M_{\rm upper}$) are calculated from the stability analysis in Section~\ref{sec:stability}.
}
\label{tab:stability}
\begin{tabular}{lcc}
\hline
Planet & $M_\mathrm{min}$ [$M_{\earth}$] & $M_\mathrm{upper}$ [$M_{\earth}$]\\
\hline
d & 0.263 & 0.66 \\
b & 0.299 & 0.75 \\
c & 0.335 & 0.84 \\
e & 0.193 & 0.48 \\
\hline
\end{tabular}
\end{table}

The above inclination constraints can be combined with injection-recovery tests to constrain the mass of undetected planets orbiting further out, for instance in the habitable zone of the star.
Injection-recovery tests in \citetalias{basant25} rule out (99\%) planets with $M\sin i > 0.57~M_{\earth}$ at the outer edge of the habitable zone.
With the minimum inclination angle of $i_\mathrm{min}=24\degr$ found above, this corresponds to an upper mass of $1.4~M_{\earth}$.
However, at the inner edge, the detection threshold is $M \sin i > 0.4~M_{\earth}$, corresponding to $M_\mathrm{upper}=0.9~M_{\earth}$.
We therefore rule out at high confidence the presence of an Earth-mass exoplanet in the inner part of Barnard's Star's habitable zone, but not the outer part.
In any case, such a hypothetical Earth-like habitable-zone planet would be well outside Barnard's Star's `abiogenesis zone' \citep[][see also Section~\ref{sec:intro}]{rimmer18}, and thus would likely lack sufficient UV flux to photochemically initiate prebiotic chemistry.

Finally, we briefly remark on the eccentricities.
When we use the eccentricities which best fit the RV data, we find the system to have a low stability probability even for $i=90\degr$, that is, when the planets' masses are the minimum masses.
This suggests that the best-fitting values of the eccentricities are unlikely to be the true values.
The posterior eccentricities from the RV data are all in fact consistent with circular orbits (within $1.5\sigma$, Table~\ref{tab:planets}; \citetalias{basant25}), so it is not implausible that the planets' orbits have simply circularised.
Another possibility is that the first three planets are in a mean-motion resonance chain; this possibility is discussed further in Section~\ref{sec:tidal_heating}.

\begin{figure*}
\includegraphics[width=\linewidth]{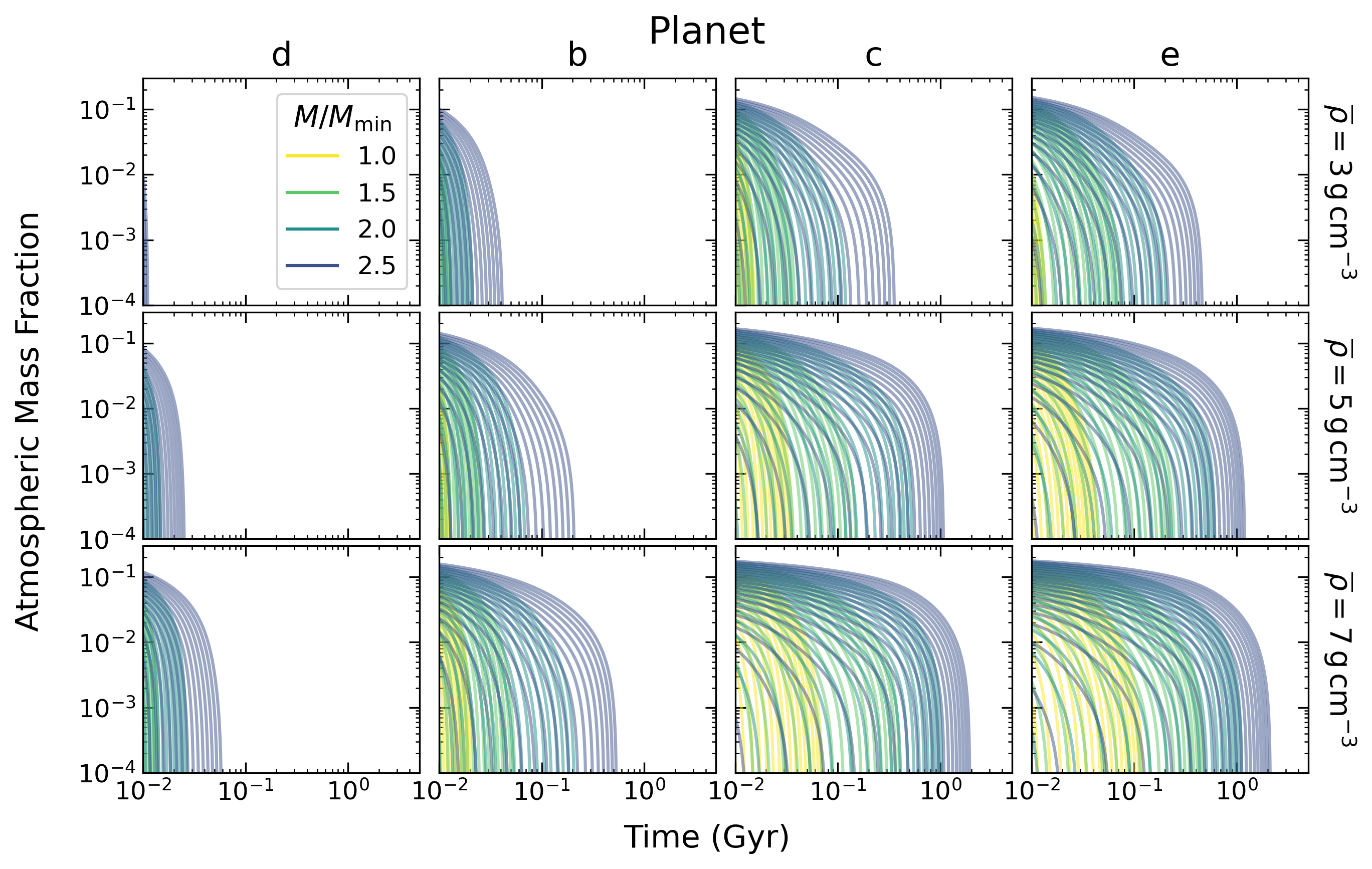}
\caption{
Atmosphere loss of Barnard's Star planets, under various assumptions of planet masses, bulk densities, and initial atmospheric mass fraction.
The mass range tested for a given planet is between its minimum mass (based on RV data; \citetalias{basant25}) and 2.5 times the minimum mass (see Section~\ref{sec:stability}).
The bulk densities $\overline{\rho}$, used to calculate planet radius in the evolution models, span a liberal range: the Solar System rocky planets have densities between $3.93~\mathrm{g\,cm}^{-3}$ (Mars) and $5.51~\mathrm{g\,cm}^{-3}$ (Earth).
No matter what assumptions are made, the planets are completely stripped of their atmospheres within $2~{\rm Gyr}$.
The atmospheres survive longer if the planets are more massive, denser, farther from the star, and begin with a larger atmospheric mass fraction.
}
\label{fig:atm_loss}
\end{figure*}

\section{Primary atmosphere loss}
\label{sec:atm_loss}

The Barnard's Star planets' low masses and tight orbits suggest that they are unlikely to have retained any primordial atmosphere inherited from the protoplanetary disc, with extreme ultraviolet (EUV) and X-ray radiation from the star having induced complete photoevaporation that the planets were not massive enough to oppose.
To verify this, we model the photoevaporative stripping of primordial H/He-dominated atmospheres, using the methodologies described in \citet{rogers21, owen20}.

The true masses, radii, and initial atmospheric mass fractions of the planets are unknown.
We therefore calculate the atmospheric evolution over a broad grid of models.
We assume that the planets
\begin{itemize}
\item have masses between 1 and 2.5 times their measured minimum masses (spanning the constraints found in Section~\ref{sec:stability});
\item have mean bulk densities $\bar{\rho}$ between 3 and $7~\text{g cm}^{-3}$;
\item initially had atmospheric mass fractions of no more than 20\%\footnote{Small planets are expected to accrete primordial atmospheres of $\lesssim20\%$ due to their shallow gravitational potential wells \citep{ginzburg16}.}.
\end{itemize}

It is clear from the simulated atmospheric evolution tracks (Fig.~\ref{fig:atm_loss}) that, even within these very liberal constraints on the planets' properties, their primordial atmospheres were photoevaporatively stripped within at most $2~{\rm Gyr}$.
The above analysis does not, however, preclude the presence of \textit{secondary}, higher-mean-molecular-weight atmospheres outgassed from the planets' mantles; the prospects for these are discussed in Section~\ref{sec:secondary_atmospheres}.

\section{Interior structure and composition}
\label{sec:interior}

The refractory composition of Barnard's Star is expected to be mirrored by its planets.
This enables inferences to be made on their interior composition and structure.
Notwithstanding several important uncertainties (discussed in Section~\ref{sec:interior_discussion}), the stellar composition alone implies an unusual mantle mineralogy for the Barnard's Star planets.

\subsection{Devolatilization modelling}
\label{sec:ggchem}

Following star formation, the protoplanetary disc cools, and elements condense out of the gas phase in decreasing order of their condensation temperatures, at which point they can be incorporated into planets.
As planet formation ends, elements with condensation temperatures below the final temperature remain mostly in the gas phase, and are relativity depleted in planets.
Condensation temperatures have long been calculated for important elements \citep[e.g.][]{wildt33, russell34, lodders03}, but it has more recently been appreciated that condensation temperatures depend significantly on the ambient composition \citep{ebel00, spaargaren25, zaveri26}.

To estimate how the protoplanetary disc of Barnard's Star would have condensed, we run the thermochemical equilibrium code \textsc{GGchem} \citep{woitke18} on Barnard's Star's composition.
Where available, we use elemental abundances from \citetalias{jahandar24}; otherwise their abundances are estimated as outlined in Appendix~\ref{app:ggchem_abunds}.

\begin{figure}
\includegraphics[width=\columnwidth]{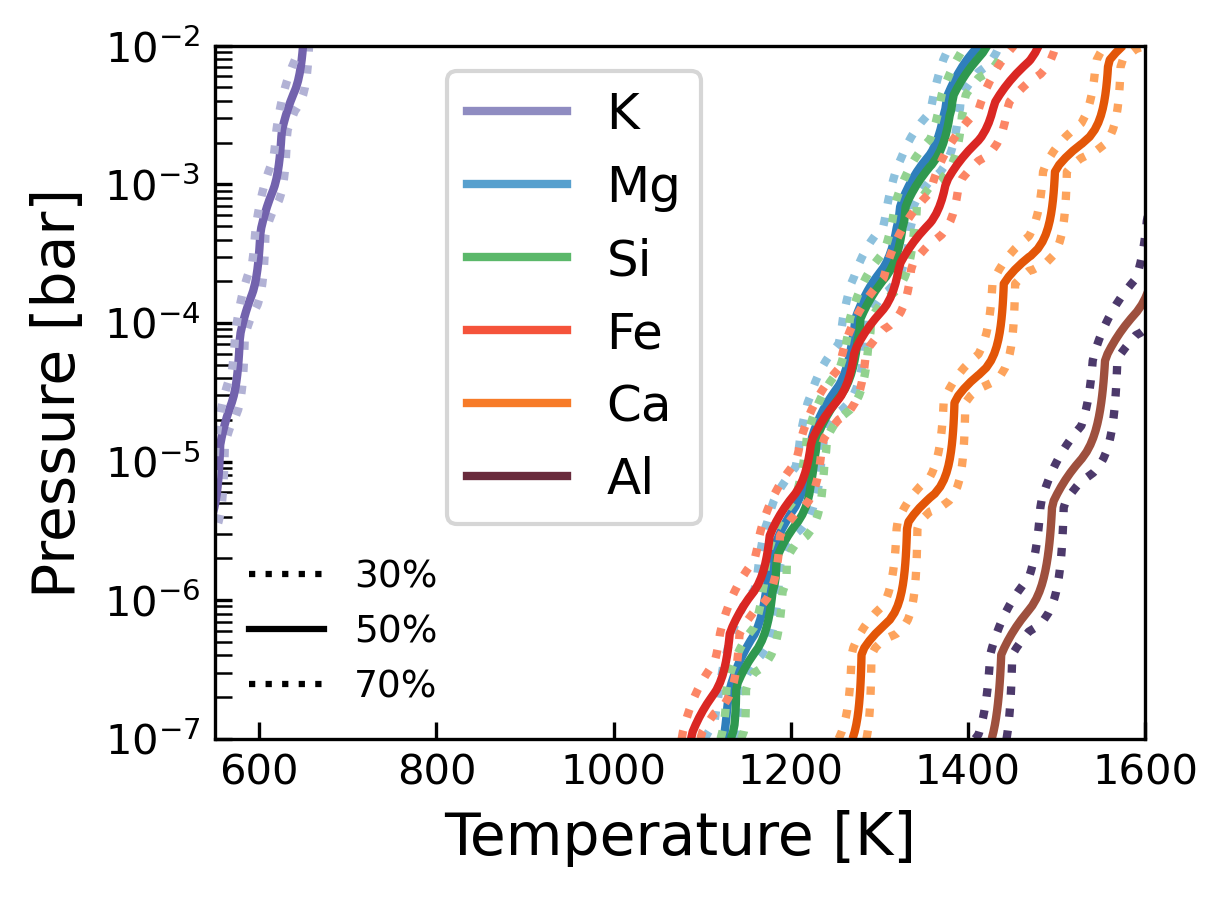}
\caption{
    Condensation of a Barnard's-Star-composition gas, calculated using \textsc{GGchem} \citep{woitke18}.
    The solid lines denote conditions where 50\% of the element is in solid form; the dotted lines denote 30\% and 70\% condensation.
    `Condensation' here is agnostic of which specific mineral phases the element is incorporated into, as long as the phase is solid.
    K is considered a moderately volatile element, condensing at the lowest temperatures; in fact the temperature of around $600~{\rm K}$ here is hundreds of degrees lower than for a Solar-System-composition gas \citep{lodders03}.
    Mg, Si, and Fe are all only slightly volatile elements, and condense at very similar temperatures.
    Ca and Al, being highly refractory elements, condense at the highest temperatures.
    The slight wiggles in the condensation lines are artefacts due to the finite grid size.
}
\label{fig:condensation}
\end{figure}

The condensation curves of six important elements are shown in Fig.~\ref{fig:condensation}.
We find condensation temperatures for each element to increase with pressure, corroborating simulations in \citet{timmermann23}.
Notably, the slopes are generally element-specific; the condensation sequence is therefore weakly pressure-dependent:
at $p=10^{-7}~\mathrm{bar}$, Fe condenses at lower temperatures than Mg and Si; at $p=10^{-3}~\mathrm{bar}$ the reverse is true.

The condensation curves of Mg and Si show that the abnormally high Mg/Si ratio of Barnard's Star is likely reflected in its planets.
The conditions at which Mg and Si condense are almost indistinguishable (Fig.~\ref{fig:condensation}), suggesting that the two elements would condense out of the protoplanetary disc and be incorporated into planets at the same time.
Barnard's Star's Mg/Si ratio is calculated in \citetalias{jahandar24} as $2.63\pm0.52$, significantly above the 95\% confidence intervals of APOGEE DR16 ($[0.87, 2.20]$; \citealt{majewski16, ahumada20}) and Hypatia ($[0.32, 1.83]$; \citealt{hinkel14}).
The fact that this abnormal abundance ratio is likely preserved in the planets has important consequences for their mineralogy \citep[e.g.,][and see Section~\ref{sec:mineralogy_models}]{delgadomena_chemical_2010, putirka_composition_2019, guimond_stars_2024}

The condensation modelling above allows for an estimate of the devolatilized composition of material from which the Barnard's Star planets are constructed.
By comparing the composition of the bulk Earth and the proto-Sun, \citet{wang19} construct a prescription for estimating bulk planetary composition, by taking the stellar composition and reducing the relative abundances of volatile elements according to their condensation temperature $T_C$.
The devolatilization factor $f_d$ by which an element with condensation temperature $T_C$ has its (linear) abundance reduced is given by:
\begin{equation}
\label{eq:devolatilization}
\log f_d = \begin{cases}
\alpha \log T_C + \beta \ \ & T_C<T_D\\
0 & T_C \geq T_D
\end{cases}
\end{equation}
where the parameters, fitted to bulk Earth and Solar abundances, take the values $\alpha=3.676$, $\beta=-11.556$, and $T_D\equiv10^{-\beta/\alpha}=1391~{\rm K}$ \citep{wang19formula}.

We now assume that the same devolatilization trend applies to the Barnard's Star planets.
This implicitly assumes that the Barnard's Star planets formed under similar conditions to the Earth; in particular, we assume that they formed \textit{in situ} inside the water snow line, and did not form with large amounts of ice before migrating inwards.
We are also assuming that the devolatilization trend is independent of the star, in the absence of stellar-mass- or metallicity-dependent extensions of this model.
These assumptions are discussed further in Section~\ref{sec:condensation_modelling}.

For each element, we extract from the \textsc{GGchem} simulations the condensation temperature at $p=10^{-4}~{\rm bar}$, and apply the above prescription to obtain devolatilized abundances.
Abundances for important rock-forming elements are shown in Table~\ref{tab:devolatilized_abundances}.
Note that we calculate Al and Ca to have $T_C>T_D$, so their abundances are not reduced from their stellar values.

\begin{table}
\caption{
    Devolatilized abundances of Barnard's Star for six important elements.
    The stellar [X/H]$_*$ values and errors are taken from \citetalias{jahandar24}.
    The condensation temperatures are calculated from our condensation modelling (Section~\ref{sec:ggchem}) as the temperature at which 50\% of an element appears in solid phase at a pressure of $10^{-4}~{\rm bar}$.
    The devolatilization factors $f_d$ are calculated according to the prescription in \citet{wang19}, given here in equation~\eqref{eq:devolatilization}.
    The devolatilized [X/H]$_p$ are shown in the final column.
    Solar abundances, where required, are taken from \citet{asplund21}.
}
\label{tab:devolatilized_abundances}
\begin{tabular}{lcrcc}
\hline
X & [X/H]$_*$ & $T_C$ [K] & $f_d$ & [X/H]$_p$ \\
\hline
K & $-0.74\pm{0.12}$ & $580$ & $0.04$ & $-2.11$ \\
Mg & $-0.33\pm{0.15}$ & $1270$ & $0.77$ & $-0.44$ \\
Si & $-0.66\pm{0.12}$ & $1280$ & $0.78$ & $-0.77$ \\
Fe & $-0.38\pm{0.03}$ & $1290$ & $0.81$ & $-0.47$ \\
Ca & $-0.60\pm{0.12}$ & $1440$ & $1$ & $-0.60$ \\
Al & $-0.40\pm{0.06}$ & $1580$ & $1$ & $-0.40$ \\
\hline
\end{tabular}
\end{table}


\subsection{Mineralogy models}
\label{sec:mineralogy_models}

\begin{figure*}
\includegraphics[width=\textwidth]{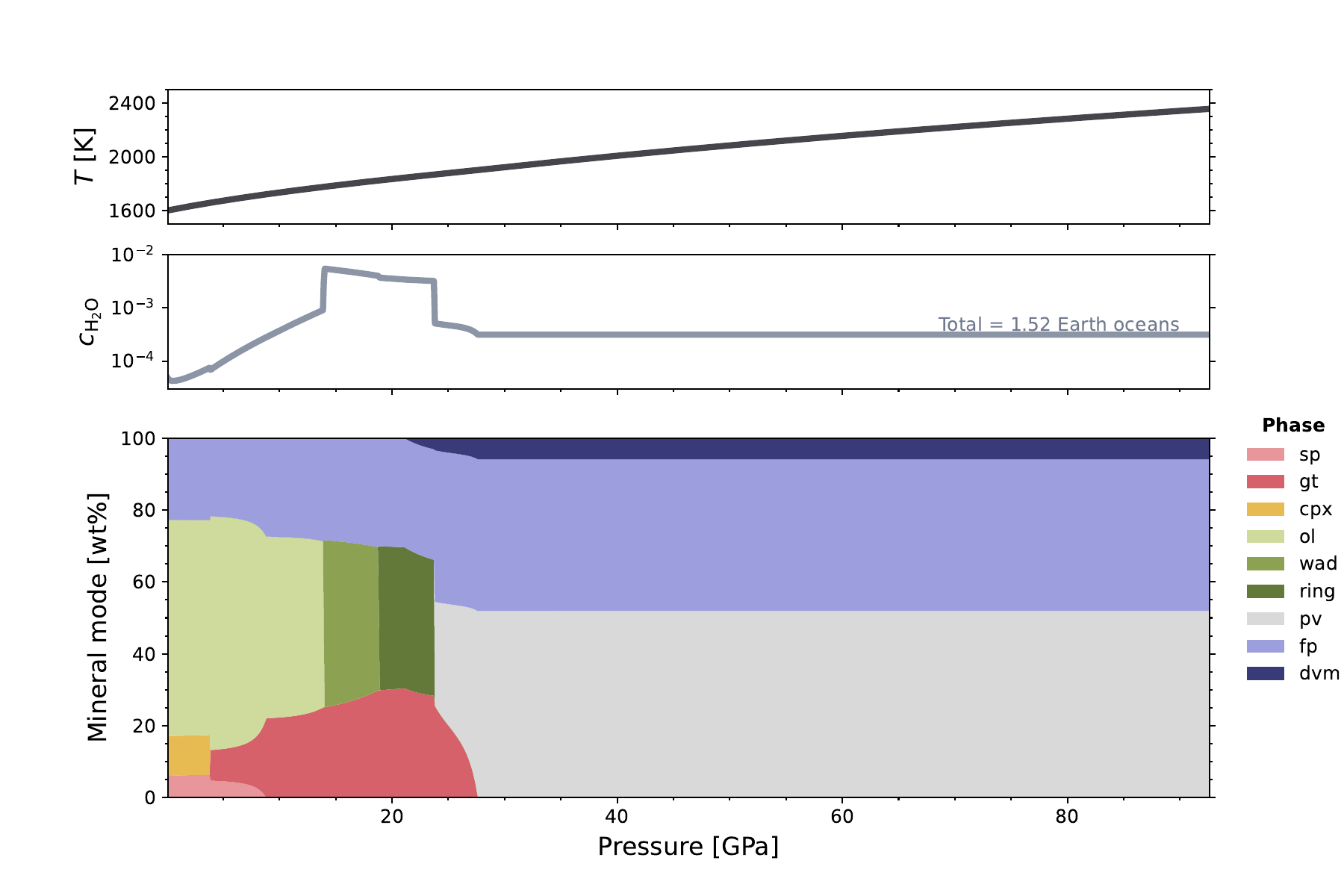}
\caption{
    The mantle mineralogy and associated water capacity estimated for Barnard's Star c, assuming an average mass of $0.59~M_{\earth}$ (Table \ref{tab:stability}) and average stellar abundances after devolatilization (Table \ref{tab:devolatilized_abundances}; Mg/Si = 2.58).
    This example calculation assumes an Earth-like distribution of bulk-planet Fe between the mantle and a metal core (corresponding to a pure-Fe core mass fraction of 32.5\% and mantl\textbf{}e FeO content of 8.2 wt\%), and a potential temperature of 1600 K.
    The core mantle boundary is at 93 GPa.
    \textit{(Bottom:)} Equilibrium mineralogy in weight percent of the mantle per pressure slice, calculated in \perplex{} \citep{connolly_algorithm_1987} using the thermodynamic database and solution model of \citet{stixrude_thermodynamics_2024}.
    Note the abundant presence of (Mg,Fe)O ferropericlase (fp) throughout the entire mantle.
    \textit{(Middle:)} The total water capacity ($c_{\rm{H_2O}}$) of the phase assemblage.
    In this example, the integral of water capacities over all pressure layers is equivalent to 1.52 Earth oceans.
    For comparison, the same model for Earth-like conditions gives just under 3 Earth oceans.
    \textit{(Top:)} The calculated mantle adiabat. 
    Mineral phases are: spinel (sp), garnet (gt), clinopyroxene (cpx), olivine (ol), wadsleyite (wad), ringwoodite (ring), perovskite (pv), ferropericlase (fp), davemaoite (dvm). 
}
\label{fig:mineralogy}
\end{figure*}

\begin{figure}
\includegraphics[width=\columnwidth]{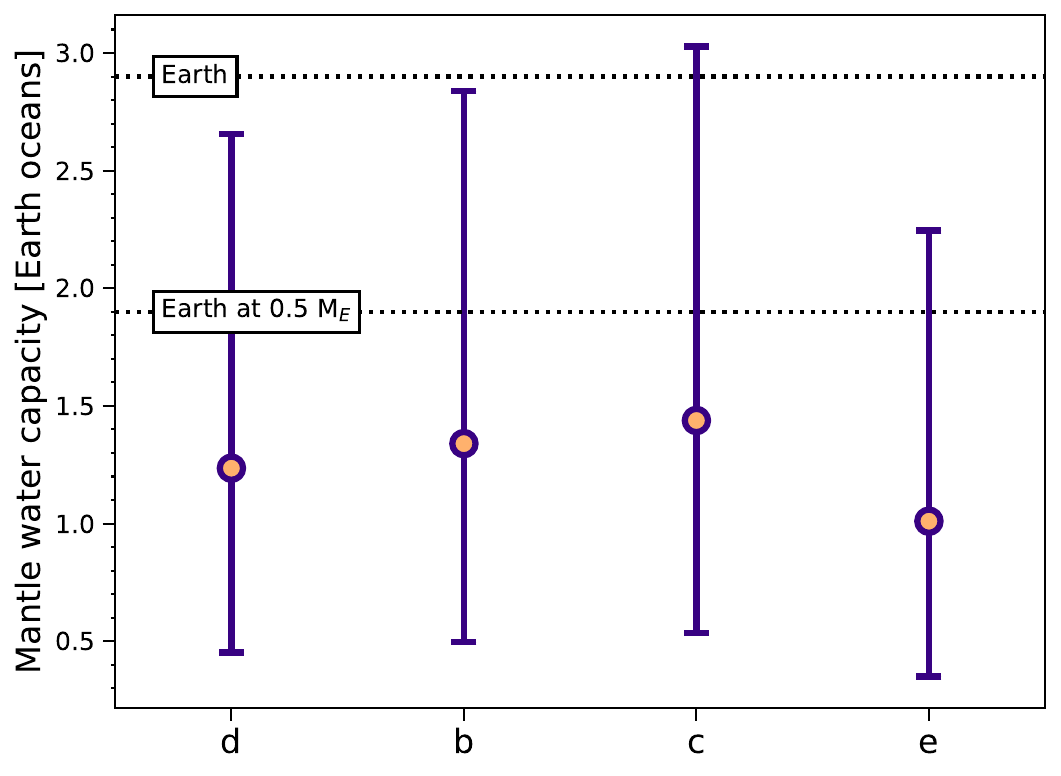}
\caption{
Mantle water capacities, in units of Earth oceans ($1.335\times10^{21}$\,kg), calculated for Barnard's Star d, b, c, and e.
For each planet, we estimate a plausible range of mantle water capacities by running the calculation at the upper and lower limits for planet mass (Table \ref{tab:planets}) and abundances for each element (Table \ref{tab:devolatilized_abundances}), and at two values of bulk-planet iron partitioning (see text).
Markers show the mean water capacity over these 48 cases per planet.
}
\label{fig:water-capacity}
\end{figure}

We use \perplex{} \citep{connolly_algorithm_1987} to estimate the mantle mineralogy of the Barnard's Star planets.
\perplex{} works by minimising Gibbs free energy to find phase abundances at equilibrium, given a thermodynamic database, a solution model, the abundances of important oxide components, and pressure-temperature ($P$-$T$) conditions.
We use the \citet{stixrude_thermodynamics_2024} thermodynamic database and solution model.
We consider the oxides \ce{SiO2}, \ce{MgO}, \ce{Al2O3}, \ce{CaO}, and \ce{FeO},  as these components make up \textgreater 99\% of the mass of Earth's mantle \citep[e.g.][]{mcdonough95}.
The abundances of these components in weight percent are determined from the devolatilized stellar abundances (Table~\ref{tab:devolatilized_abundances}); the uncertainties on these devolatilized abundances are assumed to be dominated by the stellar abundance uncertainties.
We assume that enough refractory O is available to form the first four oxides.
FeO is exceptional: planetary iron is allowed to partition between the core (as Fe-metal) and the mantle (as FeO).
This partitioning is prescribed here by the ratio $\chi^{\rm{Fe}}_m$, the molar fraction of Fe in the bulk planet which is incorporated into the mantle, which is unknown for rocky exoplanets.

Mantle $P$-$T$ is assumed to follow an adiabat, with the adiabatic gradient determined self-consistently in \perplex{}.
For all calculations we assume a mantle potential temperature of 1600 K, which does not affect mineralogy, although higher mantle temperatures would result in overall lower mantle water capacities (see later).
The planets' pressure-density-radius profiles are also determined self-consistently with mineralogy using an iterative process:
\perplex{} finds a density for each mantle layer given the equations of state in the thermodynamic model; this calculation is repeated with new $P$-$T$-dependent mineral phases until the radius of the planet surface converges for a given input mass.
The structure of the pure-Fe metal core is solved using equations of state from \citet{bouchet_initio_2013}.
These methods to calculate mantle mineralogy and interior structure self-consistently are the same as in \citet{guimond_mantle_2023}, where the methods are detailed further.


We use the resulting mantle mineral abundances to also estimate the total water capacity of the planets' mantles.
Typical mantle silicates are nominally anhydrous: \ce{H2O} is not a stoichiometric part of their chemical formulae.
However, OH-groups may still dissolve into these minerals up to their solubility limit -- the concentration of `water' in a mineral at its saturation -- which is specific to each mineral phase.
The total water capacity is the sum of each mineral phase's water concentration at water saturation, weighted by the abundance of the phase.
We use the same water solubility laws as in \citet{guimond_mantle_2023}, which build on experimental data compiled by \citet{dong_constraining_2021}; these water solubility laws are $P$- and $T$-dependent where warranted by the experimental data.
For (Mg,Fe)O, an important mineral phase in Mg-rich systems (see next paragraph), we continue to assume a constant water solubility of 10\,ppm, given the experimentally-determined solubility of water in synthetic ferropericlase of 14--79\,ppm at 25 GPa \citep{litasov_influence_2010}, and a solubility in synthetic Fe-free periclase of 3.5--9\,ppm at a lower pressure of 0.5 GPa \citep{joachim_diffusion_2013}. 
Note that we are only considering water dissolved in the mantle; we are not accounting for water from ice that could have accreted if the planets formed beyond the snow line and migrated inwards (see Section~\ref{sec:condensation_modelling}).

An representative mineralogy and water capacity calculation is shown in Fig. \ref{fig:mineralogy} for Barnard's Star c (the lower-mass planets would have truncated core-mantle boundary pressures).
The striking implication of our mineralogical modelling is that (Mg,Fe)O ferropericlase is abundant throughout the entire mantle.
On Earth, where mantle Mg/Si = 1.27, (Mg,Fe)O is the second-most abundant component of the lower mantle, but does not exist as a saturated phase in the upper mantle;
all of the MgO in the Earth's upper mantle is accommodated into \ce{(Mg,Fe)SiO3} orthopyroxene, \ce{(Mg,Fe)2SiO4} olivine, and by Ca- and Al-bearing phases.
In contrast, when Mg/Si is very high (as for Barnard's Star), all of the (Mg,Fe)O cannot be accommodated in these phases, hence it is left over as a separate phase.
\citet{spaargaren_plausible_2023} have previously demonstrated the stability of (Mg,Fe)O in upper mantles for planets with with Mg/Si ratios among the highest of those seen in the Hypatia Catalog \citep{hinkel14} when some of this bulk Si is allowed to partition into the metal core.
Even at the lower-bound Mg/Si value, (Mg,Fe)O appears above the mantle transition zone (approximately 15\,GPa for the case shown in Fig.~\ref{fig:mineralogy}); at the upper-bound it comprises $\gtrsim$60\,wt\% of the whole mantle.
Orthopyroxene never appears in our phase assemblages.
Further, although our devolatilization scheme leads to a slight enrichment of Mg/Si in the bulk planet, the minimum stellar Mg/Si ratio in Table~\ref{tab:devolatilized_abundances} still results in a stable (Mg,Fe)O phase without devolatilization. 
The presence of (Mg,Fe)O in upper mantles has consequences for the geodynamics of the Barnard's Star planets, discussed in Section~\ref{sec:thermal-evolution}, as well as for their mantles' water capacities.

The solubility of water in (Mg,Fe)O is low compared to olivine polymorphs, especially wadsleyite and ringwoodite (the two darker-green phases in Fig.~\ref{fig:mineralogy}), whose stabilities define the mantle transition zone on Earth and other silicate planets. 
Whereas wadsleyite and ringwoodite can hold up to $\sim$1\,wt\% equivalent water \citep{dong_constraining_2021}, (Mg,Fe)O can only hold $\sim$10\,ppm \citep{litasov_influence_2010}.
The consequence is that the mantle water capacities of the Barnard's Star planets are all lower than that of Earth at roughly the same mass (Fig.~\ref{fig:water-capacity}).
Whereas it is reported in \citet{guimond_mantle_2023} that water capacity generally increases with Mg/Si, there appears to be a turnover somewhere between the maximum Mg/Si value estimated therein (1.4) and that of the Barnard's Star planets (2.58).
This turnover emerges because, rather than further increasing the amounts of high-solubility wadsleyite and ringwoodite, low-solubility (Mg,Fe)O forms instead.
This effect overcomes the general trend of increasing water capacity with increasing Mg/Si reported in \citet{guimond_mantle_2023} that would otherwise be due to increasing wadsleyite and ringwoodite abundances at the expense of garnet.

The range of mantle water capacities shown in Fig.~\ref{fig:water-capacity} is calculated from the observational uncertainties in planet mass (Table~\ref{tab:planets}) and stellar abundances (Table~\ref{tab:devolatilized_abundances}), and the unknown partitioning of iron in the bulk planet, which is varied between $\chi^{\rm{Fe}}_m = [0, 0.3]$. 
For the mean elemental abundances, this range of $\chi^{\rm{Fe}}_m$ corresponds to pure-Fe core mass fractions of 37--25\% and mantle FeO contents of 0--19 wt\% respectively.
For a given mantle potential temperature, most of the uncertainty on mantle water capacity comes from varying the elemental abundances within the range in Table \ref{tab:devolatilized_abundances}; for planet c, varying the Mg/Si ratio alone contributes nearly 1 Earth ocean.
This emphasizes the importance of accurate stellar abundance estimates in the interpretation of their planets.
The mantle potential temperature itself is also important: although we have assumed a value of $1600~{\rm K}$ (based on Martian observations; \citealt{huang22}), higher potential temperatures would further destabilize ringwoodite, thus further limiting the mantle water capacity.
The sensitivity of the total mantle water capacity to certain key parameters is summarized in Table~\ref{tab:water-var}.

\begin{table}
\caption{
    Sensitivity of mantle water capacity to selected model parameters (first column).
    The rightmost column shows the range of mantle water capacities (in Earth oceans; EO) calculated for Barnard's Star c where that row's parameter is varied (in the range given in the third column), and all other parameters are set at their baseline values (second column).
    Decreasing mantle water capacities indicate a negative dependence of water capacity on that parameter.
    Symbols are:
    $M_{\rm p}$, planet mass;
    $\chi_m^{\rm Fe}$, molar ratio of iron between metal core and silicate mantle;
    $\left[{\rm X}/{\rm H}\right]_{\rm p}$, devolatilized log abundance of element X;
    $T_{\rm p}$, mantle potential temperature;
    $c_{\rm pv}$, water capacity of perovskite (relatively poorly known compared to other minerals).
}
\label{tab:water-var}
\begin{tabular}{lrcc}
\hline
Parameter & Baseline & Range & Mantle water capac. [EO] \\
 \hline
$M_{\rm p}$ [$M_{\earth}$] & $0.59$ & $[0.335, 0.838]$ & $[1.09, 1.90]$ \\
$\chi_m^{\rm Fe}$ & 0.88 & $[0.70, 1]$ & $[1.49, 1.56]$ \\
$\left[{\rm Mg}/{\rm H}\right]_{\rm p}$ & $-$0.42 & $[-0.59, -0.29]$ & $[1.86, 1.32]$ \\
$\left[{\rm Si}/{\rm H}\right]_{\rm p}$ & $-$0.75 & $[-0.89, -0.65]$ & $[1.09, 1.96]$ \\
$\left[{\rm Fe}/{\rm H}\right]_{\rm p}$ & $-$0.47 & $[-0.50, -0.44]$ & $[1.57, 1.50]$ \\
$\left[{\rm Ca}/{\rm H}\right]_{\rm p}$ & $-$0.58 & $[-0.72, -0.48]$ & $[1.49, 1.59]$ \\
$\left[{\rm Al}/{\rm H}\right]_{\rm p}$ & $-$0.40 & $[-0.46, -0.34]$ & $[1.61, 1.47]$ \\
$T_{\rm p}$ [K] & $1600$ & $[1500, 1900]$ & $[2.45, 0.57]$ \\
$c_{\rm pv}$ [ppm] & $30$ & $[30, 1000]$ & $[1.54, 2.07]$ \\
\hline
\end{tabular}
\end{table}

\section{Radiogenic heating}
\label{sec:radiogenics}

Long-lived radioactive nuclei have a strong influence on the internal dynamics of rocky planets.
These nuclei are the source of a significant (though debated) fraction of Earth's internal heat flux \citep{korenaga_urey_2008, arevalo09, kamland11, dye12, agostini20, kumaran21, abe22}.
Their abundances in planetary mantles control the convective vigour of the mantle, and hence melt production and the viability of volcanism \citep{sleep07, foley12}.
Mantle temperature also regulates heat extraction from planetary cores, and hence the viability of a geodynamo and magnetic field \citep{labrosse15, boujibar20, nimmo20}.

Barnard's Star is $7$--$12~{\rm Gyr}$ old \citep{gauza15}: much older than the Sun.
Planets born earlier in Galactic history are expected to form with higher concentrations of radioactive nuclei than younger planets.
This is because, while radioactive nuclei are synthesised in various astrophysical processes, they also decay over time; however the number of \textit{stable} nuclei monotonically increases.
A planet that forms later in Galactic history will therefore have its natal radioactive stock `diluted' by the increased abundance of stable nuclei \citep[e.g.][]{frank14}.
Conversely, the radioactive material in older planets has of course had longer to decay up to the present day.
There are thus two competing factors dictating the present-day radiogenic heat budget of a planetary mantle.
In this section we exploit the measurement of Th in Barnard's Star \citepalias{jahandar24} to estimate the radiogenic heat production in its planets, and the implications for their thermal evolution.

\subsection{Estimates of radiogenic heating from different isotopes}
\label{sec:isotopes}

Four isotopes contribute significantly to geological heat production on Gyr time-scales: $^{232}$Th, $^{238}$U, $^{40}$K, and $^{235}$U.
Each has well-characterized half-lives, specific heat production rates, and condensation temperatures (Table~\ref{tab:isotopes}).

Th and U are highly refractory elements ($T_C>1600~{\rm K}$).
Having condensed early as the protoplanetary disc cooled, exoplanets are expected to inherit their Th and U abundances from their stars.
Th and U are rarely measured directly in stellar spectra (due to low abundances and blending of lines in the optical; e.g.\ \citealt{sneden08}).
Their abundances are sometimes instead inferred indirectly by using other majority-\textit{r}-process elements as proxies, such as Eu \citep[e.g.][]{pagel89, wang20}.
In the case of Barnard's Star, 13 Th lines are identified in the infrared SPIRou spectrum, giving a well-constrained abundance measurement of $\mathrm{[Th/H]}=-0.46\pm0.04$ \citepalias{jahandar24}.
The parallel nucleosynthetic pathways of U and Th enable estimates of the present-day U abundance.
$^{40}$K abundances are more challenging, as discussed in Section~\ref{sec:K40}.

\begin{table}
\caption{
    Atomic data for the four relevant radioisotopes.
    Half-lives $\tau_{1/2}$ are from \citet{turcotte02}.
    Specific heat production rates $h$ are from \citet{ruedas17}.
    Condensation temperatures $T_C$ are from \citet{lodders03}.
}
\label{tab:isotopes}
\begin{tabular}{lcccc}
\hline
Isotope & $\tau_{1/2}$ [Gyr] & $h$ [$\mu$W~kg$^{-1}$] & $T_C$ [K]\\
\hline
$^{232}$Th & 14.0 & 26.368 & 1659 \\
$^{238}$U & 4.47 & 94.946 & 1610 \\
$^{40}$K  & 1.25 & 28.761 & 1006 \\
$^{235}$U & 0.704 & 568.402 & 1610 \\
\hline
\end{tabular}
\end{table}

To estimate the concentrations of radioactive elements in the planets' mantles, we must normalize with respect to another abundant lithophilic element, such as Mg or Si.
Despite being much more abundant than Th, their abundances are quite poorly-constrained ($\mathrm{[Mg/H]}=-0.33\pm0.15$; $\mathrm{[Si/H]}=-0.66\pm0.12$; \citetalias{jahandar24}), due to the paucity of spectral lines from these elements.
We use Mg as a normalizing element henceforth, following \citet{wang20}, but we also account for its devolatilization by a factor of 0.77 (Table~\ref{tab:devolatilized_abundances}) before its incorporation into the planets.

The remainder of this subsection quantifies the radiogenic heat production in the Barnard's Star planets' mantles, by considering each contributing radioisotope in turn.
The heat production rate $H_\mathrm{X}$ of radioisotope X, per kg of Mg, is given by:
\begin{equation}
\label{eq:heating}
H_\mathrm{X} = h_\mathrm{X}\times \Big(\frac{\mathrm{X}}{\mathrm{Mg}}\Big) \times \frac{m_\mathrm{X}}{m_\mathrm{Mg}},
\end{equation}
where $m_\mathrm{X}$ is the nuclear mass; $m_\mathrm{Mg}=24.3$ in atomic mass units.
We will frequently use the unit $\pWkgMg$, meaning `picowatt per kilogram of magnesium', to quantify specific radiogenic heat production with convenient numbers.

\subsubsection{\texorpdfstring{$^{232}$}{}Th}

Thorium is a lithophilic element, and is almost entirely sequestered in the mantles of the Barnard's Star planets; due to its very high condensation temperature (Table~\ref{tab:isotopes}) we neglect any devolatilization of Th.
Using the Th and Mg measurements from \citetalias{jahandar24} and the devolatilization of Mg (Table~\ref{tab:devolatilized_abundances}), we obtain $\mathrm{[Th/Mg]} = -0.02\pm0.16$, concluding that the Barnard's Star planets have similar concentrations of Th to the Solar System planets.
However, we note the large uncertainties, particularly on Mg, for which only three lines are observed in the original SPIRou spectrum \citepalias{jahandar24}.
Using solar abundances from \citet{asplund21}, we obtain an abundance ratio of $\mathrm{Th/Mg} = 2.90^{+1.25}_{-0.87}\times 10^{-8}$ in the Barnard's Star planets.

Th and Mg are both lithophilic elements, so we assume the Th/Mg ratio in the planets' mantles to equal the value calculated above for the bulk planet.
Equation~\eqref{eq:heating} gives a radiogenic heat production of
$H_\mathrm{Th} = 7.3^{+3.1}_{-2.2}~\pWkgMg$.

How does this compare to the Earth, and would we have found the same values had we calculated this heat production rate using solar abundances?
Using the Earth's thorium heat production rate of $Q^\mathrm{Th}=12.1^{+8.3}_{-8.6}~\mathrm{TW}$ (as measured by geoneutrino spectroscopy; \citealt{abe22}) and the Mg mass fraction of the Earth of $f_\mathrm{Mg}=0.154$ (estimated based on the composition of peridotite xenoliths; \citealt{mcdonough95}), we obtain $H_{\mathrm{Th}, \earth} = Q^\mathrm{Th} / (M_{\earth} f_\mathrm{Mg}) = (13\pm9)~\pWkgMg$.
Calculating the thorium heat production rate using solar abundances from \citet{asplund21} and devolatilizing Mg gives a heat production rate of $H_{\mathrm{Th},\earth} = 6.0^{+1.7}_{-1.3}~\pWkgMg$.
Our strategy thus gives values consistent with the observational constraints for the Sun/Earth, validating the estimation of radiogenic heat production in exoplanets from stellar abundances, and our estimation for the Barnard's Star planets of $H_\mathrm{Th} = 7.3^{+3.1}_{-2.2}~\pWkgMg$.

\subsubsection{\texorpdfstring{$^{238}$}{}U}

Effectively all the uranium in stars and planets older than a few Gyr is $^{238}$U, as its half-life is much longer than $^{235}$U (Table~\ref{tab:isotopes}).
Observationally, U is even more difficult to measure than Th, due to paucity, weakness, and blending of U~\textsc{ii} lines; indeed it is not measured in Barnard's Star in \citetalias{jahandar24}.
We instead estimate the U/Mg ratio (essentially equal to the $^{238}$U/Mg abundance) in the Barnard's Star planets by (i) interpolating the $^{238}$U/Th ratio at birth between very old stars and the Sun, (ii) estimating the relative fraction that would have decayed over the planets' lifetimes, and (iii) multiplying by the present-day Th/Mg ratio measured in the stellar spectrum.

The U/Th ratio has been measured in several evolved metal-poor stars \citep[e.g.][]{cayrel01, cowan02, frebel07}.
These stars formed early in Galactic history, leaving little time for Galactic chemical evolution (GCE) or radioactive decay to alter the $^{238}$U/Th ratio from its nucleosynthetic production ratio of $\PUTh=0.571^{+0.037}_{-0.031}$ \citep{dauphas05} before the stars formed.

While the Sun is too metal-rich for reliable spectroscopic measurements of U, Solar System abundances of $^{238}$U and Th can be measured directly in chondritic meteorites \citep{chen93}, and indirectly in the Earth using geoneutrino spectroscopy \citep{abe22}; the elements' refractory nature means that both methods should reflect the solar abundances.
Accounting for relative decay over the Solar System's lifetime, one finds that the Sun was born with a lower $^{238}$U/Th of $0.438\pm0.006$ \citep{dauphas05}\footnote{
    Note that the $^{238}$U/Th would not at formation time have been equal to the total U/Th, as there would then have been a significant quantity of $^{235}$U also.
    At the present day, though, the contribution of $^{235}$U to the numerical abundance of U is negligible, and $^{238}$U/Th$\,\approx\,$U/Th.
    Note also that Th is effectively monotopic (100\% $^{232}$Th)
}.
Evidently pre-solar GCE had reduced this ratio in the ambient interstellar medium (ISM) from its initial value of $\PUTh$ by the time the Sun formed.

Barnard's Star formed in between the formation of the Galaxy and the formation of the Sun, so we assume had an intermediate $^{238}$U/Th of $^{238}\mathrm{U/Th}=0.5\pm0.05$, in the absence of additional data.
Taking Barnard's Star to be $10~\mathrm{Gyr}$ old, we obtain a present-day $\mathrm{U/Mg} = 5.1^{+2.9}_{-1.9}\times10^{-9}$, and hence $H_{^{238}\mathrm{U}} = 4.7^{+2.7}_{-1.7}~\pWkgMg$, of the same order of magnitude as $H_{\mathrm{Th}}$.
Although $^{238}$U is only about half as abundant at formation than Th, and although more of it has decayed since birth, its contribution to the Barnard Star planets' heat budget is thus expected to be similar, due to its much greater specific heat production rate (Table~\ref{tab:isotopes}).

Using chondritic abundances from \citet{lodders09} as a proxy for solar abundances, we estimate $\mathrm{[U/Mg]}=-1.1\pm0.2$ for Barnard's Star, a depletion of a factor 13 relative to the Sun.
The metal-poor nature of Barnard's Star may mitigate the obfuscatory blending issues, enabling this abundance estimate to be tested by future observations.
Conversely, uranium abundance constraints would enable a better estimate of Barnard Star's (and its planets') age and Galactic chemical formation environment.

\subsubsection{\texorpdfstring{$^{40}$}{}K}
\label{sec:K40}

It is more difficult to accurately estimate the contribution of $^{40}$K to the radiogenic heating in Barnard's Star compared to $^{232}$Th and $^{238}$U, though it turns out to be insignificant anyway.

Firstly, K is a more volatile element than U or Th, so the link between the stellar and planetary abundances is weaker.
Indeed, K abundance varies significantly among Solar System bodies \citep{mccubbin12, kunwang21}, and is uncertain even for the Earth (e.g.\ \citealt{arevalo09}; see also \citealt{wangchen20}).
Secondly, K has two abundant stable isotopes ($^{39}$K and $^{41}$K) in addition to the radioactive $^{40}$K, which is much rarer ($\sim0.01\%$ of all K on present-day Earth).
For both of these reasons, the stellar K abundance does not inform the planetary $^{40}$K abundance.
Third, $^{40}$K is sourced by multiple nucleosynthetic pathways \citep{clayton03, the07, kunwang21}, making the identification of proxies much more difficult than for Eu and pure-\textit{r}-process Th and U.

We are forced to use theoretical GCE models to estimate the abundance in Barnard's Star.
At a Galactic age of $3~\mathrm{Gyr}$ (corresponding roughly to Barnard's Star's formation time), \citet{frank14} give an ISM mass fraction of around $2.5\times 10^{-4}$ for Mg (their fig.~3); for $^{40}$K a value of around $2\times10^{-9}$ is given, with an uncertainty of about a factor of 2 depending on the contributions of different nucleosynthetic pathways (their fig.~5).
Accounting for isotope masses and relative devolatilization (Table~\ref{tab:devolatilized_abundances}), this corresponds to a $^{40}$K/Mg ratio of $2.7\times10^{-7}$ at formation.
After $10~\mathrm{Gyr}$ of radioactive decay, this falls to around $1.0\times10^{-9}$.
We then arrive at a heat production rate of $H_{^{40}{\rm K}}=0.05~\pWkgMg$, to within a factor of 2.
Even with large uncertainties, the contribution of $^{40}$K to the planets' present day heat budget is thus expected to be subdominant to $^{232}$Th and $^{238}$U.
This is largely a consequence of the isotope's shorter half-life (Table~\ref{tab:isotopes}).
However, $^{40}$K could have contributed significantly to the earlier-stage evolution of the planets.

\subsubsection{Total}

The radiogenic heat contributions at present day from $^{232}$Th, $^{238}$U and $^{40}$K in the Barnard's Star planets were calculated above to be $7.3^{+3.1}_{-2.2}~\pWkgMg$, $4.7^{+2.7}_{-1.7}~\pWkgMg$, and negligible respectively, for a total of $H_{\rm BS} = (12\pm3)~\pWkgMg$.
Other isotopes are not expected to contribute significantly today.
For comparison, Earth's total radiogenic heat production is estimated to be $Q\approx22~\mathrm{TW}$ (\citealt{oneill20}; but see also \citealt{agostini20}), corresponding to $H_{\earth} = Q / (M_{\earth} f_\mathrm{Mg}) = 24~\pWkgMg$.
The Barnard's Star planets' mantles are thus expected to be generating about half as much specific radiogenic heat as Earth today, primarily due to the greater quantity of $^{238}$U having decayed away since formation much longer ago.

\subsection{Tidal heating: tidally locked, but a potential mean-motion resonance chain}
\label{sec:tidal_heating}

In addition to radiogenic heating, another potential internal heat source for the planets is tidal heating.
We consider here two phenomena which dictate the importance of tidal heating: tidal locking, and eccentricities.

The Barnard's Star planets are likely tidally locked.
For the TRAPPIST-1 system, \citet{turbet18} calculate tidal evolution time-scales for the planets' rotations to be between $10^{-4}~{\rm Myr}$ for TRAPPIST-1b ($a=0.012~{\rm au}$) and $7~{\rm Myr}$ for TRAPPIST-1h ($a=0.062~{\rm au}$).
Theoretical calculations \citep{gladman95} suggest that the tidal locking time-scale $\tau_{\rm TL}$ scales as:
\begin{equation}
\label{eq:locking}
\tau_{\rm TL} \propto \frac{a^6 I_p Q}{M_*^2 k_2 R_p^5} \propto \frac{\alpha Q \rho_p}{k_2} \frac{a^6}{M_*^2},
\end{equation}
where $I_p=\alpha M_p R_p^2$ is the planet's moment of inertia, $Q$ is its specific dissipation function, $k_2$ is its tidal Love number, and $\rho_p$ is its bulk density.
The quantities $\alpha$, $Q$, $k_2$, and $\rho_p$ are largely `intensive' functions of the planet's interior structure, and given that the Barnard's Star planets and TRAPPIST-1 planets are all small and rocky, we will assume these quantities to be of the same order of magnitude for each.
The tidal locking time-scale thus scales approximately with $a^6/M_*^2$.
Accounting for the respective stellar masses and orbital radii, the tidal locking time-scales for the Barnard's Star planets are calculated to be of order $10^6$--$10^7~{\rm yr}$.
As Barnard's Star is $\sim10^{10}~{\rm yr}$ old \citep{gauza15}, the planets likely became tidally locked long ago.
The planets' rotations are therefore not expected to lead to significant tidal heating.

Tidal heating can also occur as energy is dissipated during the circularization of initially eccentric orbits.
The TRAPPIST-1 planets have eccentricities between 0.002 and 0.01, at between $2\sigma$ and $15\sigma$ confidence \citep{grimm18}; these eccentricities are maintained by a mean-motion resonance chain \citep{luger17}, and facilitate a degree of tidal heating that is expected to outweigh radiogenic heating \citep{barr18, dobos19}.
The Barnard's Star planets' eccentricities are less well-constrained by the available RV data: the best-fitting values are between 0.03 and 0.08, above zero at only around $1.5\sigma$ significance \citepalias{basant25}.
We have found, however, that the best-fitting eccentricities lead to unstable orbits under certain assumptions (Section~\ref{sec:stability}), whereas models with circular orbits are stable for a range of system inclinations (i.e.\ planet masses).
We cannot however rule out small, non-zero eccentricities maintained by a mean-motion resonance chain, as occurs in the TRAPPIST-1 system.

The inner three planets in the Barnard's Star system (d, b, c) have period ratios that are both close to the 4:3 resonance -- $(P_b/P_d, P_c/P_b)=(1.348, 1.308)$ -- raising the question of whether a mean-motion resonance chain is in operation.
Such a chain could sustain non-zero eccentricities in the system, leading to significant tidal heating.
It is challenging to confirm or rule out a resonance chain through $N$-body simulations, due to uncertainties in the planets' orbital parameters, particularly mean longitudes.
For transiting systems, the transit times readily provide mean longitudes, enabling $N$-body-simulation-based checks for resonances \citep[e.g.][]{macdonald22, quinn23, lammers24}.
However, RV data are typically phase folded to enhance the signal size, and as a result the mean longitudes are often poorly-constrained.
Indeed this is the case for the Barnard's Star system, where the arguments of periapsis are largely unconstrained \citepalias{basant25}.
We are therefore not able to conduct reliable $N$-body simulations on the Barnard's Star system without a full sweep of mean longitudes, arguments of periapsis, and eccentricities for all four planets, which we deem beyond the scope of this work.
Further RV observations of Barnard's Star would better constrain the planets' eccentricities and mean longitudes, enabling more detailed analyses of the roles of resonances and eccentricity-induced tidal heating.

\subsection{Thermal evolution}
\label{sec:thermal-evolution}

To connect the threads presented throughout this work, we take a first step in modelling the long-term evolution of this old, Mg-rich, Th-poor planetary system.
We estimate how the present-day planetary interior thermal states would play out with
(i) our estimated interior structures,
(ii) our estimated radiogenic element abundances, and
(iii) the possible viscosities of ferropericlase-rich mantles as inferred from the literature.
We do however neglect the unknown contribution of tidal heating (Section~\ref{sec:tidal_heating}).
The following section shows how we can capture the evolution with a simple 0D model. 

In models of rocky exoplanet thermal evolution, the widespread flow laws for Earth-like mineralogies are usually assumed, in the absence of any compositional information suggesting otherwise.
However, Mg-rich or -poor mantles are expected to have different rheological behaviour to pyrolitic compositions \citep{ballmer17, spaargaren20}.
We must therefore account for our inference of abundant mantle ferropericlase when exploring the thermal evolution of these planets. 

Solid-state convective heat transport within a planetary mantle is largely controlled by the deformation of the weakest (sufficiently abundant) mineral phase.
Whether ferropericlase is weaker or stronger than bridgmanite is debated \citep[e.g.,][]{cordier23}, although the `strong ferropericlase' scenario seems to require crystal deformation mechanisms which are hard to reconcile with geophysical observations of the Earth's mantle \citep[and references therein]{karato_rheology_2025}.
Interestingly, whilst a weaker ferropericlase viscosity complicates our view of Earth's lower mantle because ferropericlase is less abundant than bridgmanite -- so we would need to determine how abundant ferropericlase must be for it to dictate how Earth flows -- ferropericlase being at least as abundant as bridgmanite in lower mantles of Barnard's-Star-like composition (Fig. \ref{fig:mineralogy}) erases this complication, because this weak and abundant phase is almost inevitably the main controller of the viscosity.

\begin{figure}
\includegraphics[width=0.99\columnwidth]{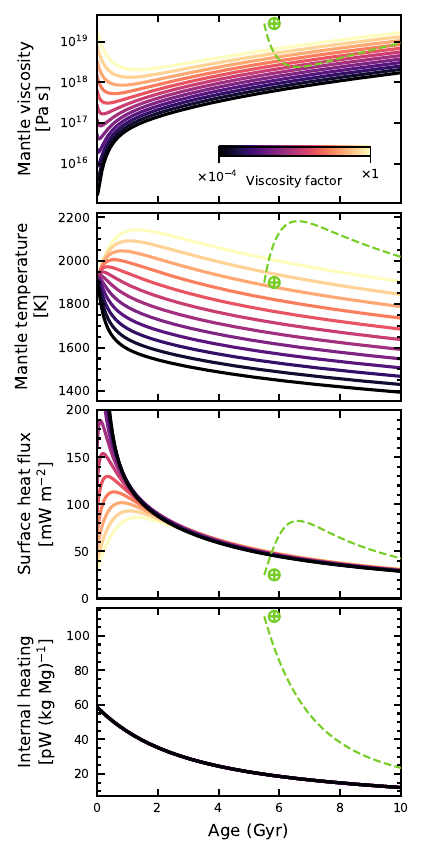}
\caption{
Simple thermal evolution of a Barnard's-Star-c-like planet ($0.59~M_{\earth}$) in the stagnant lid regime, with interior structure from Fig.~\ref{fig:mineralogy} and radiogenic element abundances from Section~\ref{sec:isotopes}.
From top to bottom, we show tracks of:
mantle viscosity;
0D mantle temperature (corresponding to a representative temperature below the stagnant lid, higher than the potential temperature);
the heat flux out of the surface;
and the internal heating rate due to radiogenic element decay.
Coloured lines show cases where temperature-dependent viscosity is scaled by a prefactor ($\eta_0$ in equation~\ref{eq:viscosity}) with respect to dry olivine \citep{karato_rheology_1993}, demonstrating possible effects of a high-Mg/Si mineralogy.
The green dashed lines show the same model applied to an Earth-like composition at $0.59~M_{\earth}$, for comparison.
While the surface heat flux adjusts to accommodate the internal heating rate, larger viscosity prefactors mean that the mantle needs to be hotter to maintain the same viscosity, and vice versa.
}
\label{fig:thermal-evolution}
\end{figure}

Fig.~\ref{fig:thermal-evolution} applies the parameterised thermal history model detailed in \citet{guimond_blue_2022} to a model planet with Barnard's-Star-c-like quantities:
a planet mass with the mean value of $0.59~M_{\earth}$;
planet and core radii and average densities from our interior structure model (Section \ref{sec:interior}); an assumed age of $10~{\rm Gyr}$;
and radiogenic element abundances from Section \ref{sec:isotopes}.
(Note, however, that the backpropagation of internal heating only accounts for $^{238}$U and Th, and does not include the poorly-constrained early-time contributions of $^{40}$K, $^{235}$U, and shorter-lived isotopes.)
The thermal history model does not consider water cycling, melting, and their associated feedbacks, leaving such detailed modelling and corresponding assessment of the additional unknown parameters to future work.
We show models for the most massive planet only; the smaller planets in the system are expected to be colder, all else equal, due to their larger surface-area-to-volume ratio \citep[e.g.][]{stevenson_styles_2003}. 

To bracket the mineralogical effect of rheology, Fig.~\ref{fig:thermal-evolution} shows thermal evolution tracks where the mantle viscosity,
\begin{equation}
\label{eq:viscosity}
\eta(T_m) = \eta_0 A \exp\left(\frac{E_a}{R_b T_m}\right),
\end{equation}
is scaled by an unknown factor, $\eta_0 \in [10^{-4}, 1]$.
In equation~\eqref{eq:viscosity}, $R_b$ is the universal gas constant, $T_m$ is mantle temperature, $E_a$ is the activation energy (approximately $300~{\rm kJ~mol}^{-1}$ for both olivine and ferropericlase; \citealt{karato_rheology_1993, yamazaki_mineral_2001}), and $A=1.6\times10^{11}~{\rm Pa~s}$ is another factor incorporating grain size and other fixed material properties.
The minimum $\eta_0$ corresponds to the approximate difference between the diffusion creep of dry olivine from \citet{karato_rheology_1993} and the diffusion creep of ferropericlase from \citet{yamazaki_mineral_2001} at 1873\,K and 25\,GPa.
We expect that allowing $\eta_0$ to vary in this way will capture the plausible range of viscosities at a given temperature for aggregate mantles with high volumes of ferropericlase, without attempting to prescribe exactly how viscosity should depend on the phase assemblage, which remains uncertain.

Fig.~\ref{fig:thermal-evolution} demonstrates how the mantle viscosity law controls mantle temperatures when viscosity decreases with increasing temperature.
The heat flux out of the surface (third panel) quickly adjusts to evacuate the steadily-decreasing heat generated internally (fourth panel), such that the mantle does not heat up catastrophically.
There will be some convective viscosity required to transport this internal heat to the surface (first panel). The required viscosity has a pre-determined temperature, per equation~\eqref{eq:viscosity}.
For overall-weaker viscosities (lower $\eta_0$), temperatures do not need to be as high to achieve this required viscosity (second panel).
The combination of low internal heating and weak viscosity leads to cold mantles.
The lower mantle temperature has consequences for the viability of volcanism, outgassing, and the presence of secondary atmospheres, as discussed in Section~\ref{sec:secondary_atmospheres}.

\section{Discussion}
\label{sec:discussion}

The preceding sections have investigated a variety of aspects of the Barnard's Star planetary system, and the nature of the individual planets.
This section discusses some implications of these analyses.

\subsection{Prevalence of Lilliputian planetary systems}

The mass constraints derived in Section~\ref{sec:stability_results} conclude that even the most massive planet in the system (c) is very likely less massive than the Earth; all four planets are hence in between the masses of Earth and Mars ($0.11~M_{\earth}$).
This curious architecture raises questions about the formation and evolution of the system.

To date only two other planetary systems have been found with 4+ exoplanets with sub-Earth mass, minimum mass, or radius\footnote{
    As found in \url{https://exoplanet.eu/catalog}; accessed Jun 2025.
}: Kepler-444~A \citep{campante15}, and Kepler-1542~A \citep{morton16}.
We dub these `Lilliputian' systems\footnote{
    This term is a reference to the satirical novel \textit{Gulliver's Travels}, in which the fictional island of Lilliput is inhabited by tiny people.
}.
Population-level studies on such planetary systems are not yet permitted by observational constraints.
However, these three systems present interesting avenues of further investigation as these statistics become available in the future.

\subsubsection{Metallicity}

Barnard's Star's hosting of several small planets is not reliably attributable to its low metallicity.
While there is a strong metallicity-occurrence relationship for gas giant planets \citep[e.g.,][]{fischer05}, this relationship weakens for Neptune-sized planets \citep[e.g.,][]{sousa11} and vanishes for terrestrial planets \citep{buchhave12}.
Kepler-444~A and Kepler-1542~A have estimated metallicities of respectively $-0.55$ \citep{campante15} and $0.06$ \citep{morton16}, bracketing Barnard's Star's $-0.38$ \citepalias{jahandar24}.
These two stars indicate that the presence of multiple small planets could be similarly independent of metallicity.

\subsubsection{Stellar mass}

Lilliputian systems could be limited to low-mass single stars, or larger stars in multi-star systems.
General exoplanet occurrence has been shown to correlate with stellar mass \citep[e.g.,][]{lovis07, johnson10}, and giant planets have long been shown to be rarer around M dwarfs \citep[e.g.][]{endl06}, likely due to larger stars hosting more dust-massive protoplanetary discs \citep{pascucci16}.
Kepler-444~A ($0.76~M_{\sun}$; \citealt{campante15}) and Kepler-1542~A ($0.94~M_{\sun}$; \citealt{morton16}) are both much more massive than Barnard's Star ($0.16~M_{\sun}$).
However, the protoplanetary discs of the larger stars may have been truncated by the presence of M dwarf companions Kepler-444~BC and Kepler-1542~B, reducing the amount of material available for planet formation in each case \citep{wang14, dupuy16, sullivan23}.
This potentially stunted the growth of its planets, which perhaps would have grown larger in the absence of the companions.
The system of eight planets, including three sub-Earths, orbiting TRAPPIST-1 ($0.089~M_{\sun}$, \citealt{delrez18}; \citealt{agol21}) and the system of three sub-Earths orbiting LHS~1678 ($0.35M_{\sun}$; \citealt{silverstein22}) further hint at Lilliputian systems being more common around low-mass stars.

However, it is possible that stellar mass only appears to be important for hosting multiple small planets due to observational bias.
Small planets are of course more difficult to detect around larger stars (regardless of whether they are transiting, or detected by RV).
It is therefore plausible that many multi-sub-Earth systems exist around larger stars, but have escaped detection.
Indeed, the proximity of Barnard's Star may exert an additional bias, given that planets are also generally easier to detect around nearer stars.
The discovery of more such systems, likely to emerge from the PLATO mission \citep{plato}, is needed to concretely assess their relationships to stellar mass and metallicity.

\subsubsection{Outer giant planets}

The ratios of adjacent planets' orbits hint at a lack of outer giant planets in the system.
The gap complexity $\mathcal{C}$ of the system -- a measure of the unevenness of the planets' orbital periods in log space; $\mathcal{C}=0$ corresponds to planets with identical period ratios \citep{gilbert20} -- is $0.131$, somewhat higher than the median of \textit{Kepler} systems ($0.10$; \citealt{he23}), and similar to that of the inner Solar System ($0.126$; \citealt{rice24}).
The presence of outer giant planets has found to correlate with high ($\gtrsim 0.3$) gap complexity \citep{he23}, suggesting that such planets are unlikely to be present in this system; they are furthermore ruled out by injection-recovery tests on the RV data for periods $\lesssim10^3~{\rm days}$ \citepalias{basant25}.

\subsection{Low prospects for atmospheres}

The atmospheric evolution simulations in Section~\ref{sec:atm_loss} show that a primary atmosphere could last at most $2~{\rm Gyr}$ on any of the Barnard's Star planets.
In fact, the only simulations in which a primary atmosphere lasted more than a Gyr were those for which the planets had the highest masses ($M/M_{\rm min}=2.5$), or improbably high bulk densities ($\bar{\rho}=7~{\rm g}~{\rm cm}^3$).
Most likely, a primary atmosphere could survive for of order $0.1~{\rm Gyr}$ on the outer planets c and e, and of order $0.01~{\rm Gyr}$ for the inner planets d and b.

\subsubsection{Secondary atmospheres}
\label{sec:secondary_atmospheres}

The radiation environment and interior water storage capacities of the Barnard's Star planets limit their prospects for hosting secondary atmospheres.
While primary atmospheres are limited to the first $\lesssim2~{\rm Gyr}$ in the Barnard's Star planetary system (Fig.~\ref{fig:atm_loss}), secondary atmospheres could in principle be outgassed later in a planet's lifetime \citep[e.g.][]{jakosky25}.
However, from the modelling in this work, there emerge multiple arguments against the presence of secondary atmospheres as well.

First, the water storage capacity of the Barnard's Star planets' mantles is predicted to be significantly lower than that of the Earth (Fig.~\ref{fig:water-capacity}).
This suggests that, even if the planets had formed with their mantles entirely saturated with water, the total inventory of water available for later outgassing would still be relatively low.
Further, this water capacity need not be taken up: the mantle minerals could have always been dry.

Second, secondary atmospheres are subject to many of the same escape phenomena as primary atmospheres, which escape quickly as shown above.
It is true that secondary atmospheres have a higher mean molecular weight (and, as a result, smaller scale height) than H/He primary atmospheres, and that they may only be present at late times when the star emits less EUV flux.
However, secondary atmospheres are also much less massive overall, having been outgassed from minor interior sources, rather than collected from gas in the ambient protoplanetary disc.
Additionally, despite its old age Barnard's Star continues to flare occasionally \citep{paulson06, france20}, further threatening any extant atmospheres.
Detailed atmospheric escape modelling, including higher-order effects such as sputtering not considered here, for a theoretical Mars-mass but further-out planet orbiting Barnard's Star is carried out in \citet{brain26}, where it is found that a secondary atmosphere is unlikely to survive the considerable EUV flux.

Third, even if the planets' mantles had retained water at the saturation limit of $\sim$1 Earth ocean, and if present-day atmosphere escape rates were low, we infer that efficient volcanism and outgassing are unlikely due to relatively low mantle temperatures and heat fluxes (Fig.~\ref{fig:thermal-evolution}).
It was shown in Section~\ref{sec:isotopes} that the planets are expected to have a lower concentration of radioactive material than the Earth, owing to their greater age.
An additional effect is the planets' smaller size: while total mantle heat production scales with a planet's volume, the heat flux scales with the planet's surface area.
The higher surface-area-to-volume ratio for the Barnard's Star planets exacerbates heat loss compared to the Earth, implying cooler mantles at the present day.
Together with viscosity effects (Fig.~\ref{fig:thermal-evolution}), these effects limit the propensity for volcanism, which would otherwise enable the release of stored volatiles into the atmosphere.
However, we note that the thermal evolution modelling in Section~\ref{sec:thermal-evolution} does not incorporate melting, volcanism, or outgassing, as this would require much more detailed geodynamic modelling.
Nor does our modelling involve tidal effects, which may also be important (Section~\ref{sec:tidal_heating}).
Although well beyond the scope of this work, these extensions could better constrain the geodynamic state and outgassing history of these planets.

Due to these factors, neither primary nor secondary atmospheres are expected on any of the Barnard's Star planets.

\subsubsection{Atmospheric observation}

Our hypothesis that none of the four known planets in the Barnard's Star system has an atmosphere will be testable with the upcoming European Extremely Large Telescope (ELT).
Facilities such as ELT will target the planetary reflection spectra, since these worlds do not transit and are unlikely to have significant thermal emission at wavelengths ($<5\micron$) typically observed from the ground.
These reflection spectra contain information on the scattering properties of the surface and atmosphere.
Therefore, measuring them may reveal whether an atmosphere is present.

To measure the reflected light spectrum of these worlds with the ELT, the planets must be spatially resolvable from their host star and bright enough that their spectrum can be extracted from the stellar glare.
At maximum separation, these planets lie within approximately $20~{\rm mas}$ of their star, which is similar to the diffraction limit of the ELT at the wavelengths at which the METIS instrument observes.
Therefore, METIS/ELT will not be able to measure the spectra of these planets as it is unable to spatially resolve them.
They are spatially resolvable with HARMONI/ELT, but the high-contrast mode of this instrument employs a focal plane mask $30~{\rm mas}$ in radius.
While offsetting the mask may enable observations at smaller separations
\citep{vaughan24}, it is likely that the planets will be too faint for their spectra to be extracted in a reasonable amount of observing time.
ANDES/ELT on the other hand \textit{will} be able to spatially resolve these planets, and is already planning to characterize other planets at similar separations and with similar star-planet contrast ratios as part of its golden sample \citep{marconi24}.
It is therefore possible that ANDES/ELT could be used to measure the reflection spectra of these worlds and reveal whether they have atmospheres.

\subsection{Interior modelling caveats}
\label{sec:interior_discussion}

The conversion of a set of stellar abundances into a planetary interior mineralogy involves a number of assumptions, most of which are reviewed in \citet{guimond_stars_2024} and \citet{wang22b}.
Further caveats specific to the analysis conducted in this work are presented here.

\subsubsection{Condensation modelling}
\label{sec:condensation_modelling}

An important implicit assumption made in this work is that the Barnard's Star planets did not form beyond the water snow line, before later migrating inwards to their current locations.
If they had formed further out, then they would have contained a significant quantity of ice at formation; this may have facilitated complex aqueous reactions with the rock, altering the mineralogy of the mantle and core in ways that would be difficult to model.
Furthermore, the accretion of large quantities of ice could of course increase the amount of water present somewhere in the planets beyond what could be dissolved in the mantle: extra water could have gone into initial steam atmospheres, for example.
This possibility is difficult to rule out, and cold formation remains a possible scenario.
Notwithstanding uncertainties regarding the volatile composition, the \textit{refractory} composition estimated in this work -- and hence the water dissolved in the mantle -- remains reliable.
For example, the condensation conditions of Mg and Si are so concomitant (Fig.~\ref{fig:condensation}) that the high Mg/Si ratio of Barnard's Star is likely to be replicated in the planets even if the formation conditions were quite different.

Another assumption implicit in our use of \textsc{GGchem} in modelling the condensation of a Barnard's-Star-composition primordial gas, is the assumption of chemical and phase equilibrium.
While fast reaction rates are expected to beget equilibrium quickly at the temperatures relevant to a protoplanetary disc, non-equilibrium processes may undermine this assumption.
The formation of grains of size greater than $\sim0.01~{\mu{\rm m}}$ shields the interior of the grains from the disc, locking the inside material out of equilibrium.
In high-C/O discs, some processes (particularly hydrogenations; \citealt{lewis80}) are not fast enough even at protoplanetary disc temperatures to reach equilibrium before the disc disperses \citep{spaargaren25}.
However, as Barnard's Star's C/O ratio is subsolar ($0.39\pm0.32$; \citetalias{jahandar24}), such slow processes are not expected to significantly affect the condensation modelling.
This low C/O ratio has also avoided the significant deviations of refractory element ratios from stellar values expected in highly-reduced discs \citep{zaveri26}.

We also mention that although \textsc{GGchem} accounts for a vast number of gas and mineral species, this does not include solid solutions, which are only represented as combinations of endmembers \citep{woitke18}.
The relative stability of solid solutions, compared to mixtures of pure endmembers, may slightly alter the condensation temperatures derived in Section~\ref{sec:ggchem}.

Following condensation, there are also caveats to the calculation of the devolatilized abundances (Section~\ref{sec:ggchem}; Table~\ref{tab:devolatilized_abundances}).
The condensation temperatures were calculated at a pressure of $10^{-4}~{\rm bar}$, but it is shown in Fig.~\ref{fig:condensation} that these temperatures are weakly pressure-dependent, and the pressure at which planetesimals form in a protoplanetary disk is not agreed upon in modelling work (e.g., $4\times 10^{-5}~{\rm bar}$, \citealt{cameron95}; $10^{-5}~{\rm bar}$, \citealt{willacy98}), nor in Solar System observations (e.g.\ \citealt{kornacki86}).
However, within an order of magnitude of uncertainty in the pressure, the uncertainties in the condensation temperatures are minor.
In any case, the most consequential ratio for the composition of the Barnard's Star planets' mantles is Mg/Si, and both elements condense at almost exactly the same temperature regardless of pressure.
This ratio -- and the resulting plenitude of ferropericlase -- is therefore unlikely to be affected by the exact pressure chosen.

A more important caveat may be that the devolatilization prescription used here (equation~\ref{eq:devolatilization}) is derived for the Earth's relation to protosolar abundances \citep{wang19formula}.
This prescription may be invalid here: the Barnard's Star planets orbit a very different star, and at a much closer orbital radius.
However, we note that even significant variations in the devolatilization scaling have been shown to produce qualitatively similar interior structures \citep[][their section 4.1]{wang22b}.
More detailed modelling of condensation in protoplanetary disks of non-Sun-like stars will confirm how universal the Earth-Sun prescription is, and hence how valid its application is here.

\subsubsection{Mineralogical uncertainties}

A caveat to the mineralogical calculations is that silicate mineral phase equilibria are not constrained experimentally at Mg/Si ratios as high as those seen in Barnard's Star.
Estimating the mineralogy of the Barnard's Star planets using \perplex{} therefore requires extrapolation beyond the limits of thermodynamic databases.
The presence of a free MgO phase is dictated stoichiometrically -- there is too much Mg to be taken up by Si-bearing minerals -- but the true phase assemblage remains to be tested none the less.

Further, we have assumed that enough refractory O is available to form these silicate phases in the first place.
The total amount of O accreted by rocky planets is difficult to estimate from disk condensation modelling.
It is possible that if the Barnard's Star planets formed in an O-poor environment, then excess Mg would remain as metallic Mg, rather than MgO, and likely enter the planets' metallic cores.
We note however that stellar abundances of Barnard's Star show a Mg/O ratio consistent with the Sun's (${\rm [Mg/O]}=0.08\pm0.15$; \citetalias{jahandar24}).

Lastly, our water capacity estimates assume a fixed mantle potential temperature of $1600~{\rm K}$, which is generally appropriate for small rocky planets \citep[e.g.][]{huang22}.
Higher mantle potential temperatures would lead to lower water capacities overall (see Table~\ref{tab:water-var}; also \citealt{guimond_mantle_2023}).
However, the results of our thermal evolution modelling suggest that much hotter mantles are less likely.

\subsubsection{Post-formation processing}

Finally, we note that there are some events that can dramatically alter the bulk composition of exoplanets after formation, which cannot be accounted for with the methods used here.
Mercury's bulk composition is not reflective of the solar composition, its mantle having potentially been blasted off by a giant impact in its early history \citep{benz07}.
Indeed, subtle aspects of Earth's non-chondritic composition have been proposed to be a consequence of collisional processing early in its accretion history \citep[e.g.][]{bonsor15}.

\section{Conclusions}
\label{sec:conclusion}

This work presents an analysis on several fronts of the Barnard's Star planetary system, the closest planetary system of a single star \citepalias{basant25}.
Under the assumption of dynamical stability, we find that each planet is very unlikely to be more than 2.5 times its minimum mass.
Barnard's Star therefore hosts a `Lilliputian' system, with all four planets between the masses of Earth and Mars.
By devolatilizing the elemental abundances measured in Barnard's Star \citepalias{jahandar24}, we model the planets' mantle mineralogies.
The extremely high Mg/Si ratio of the star (2.63) is approximately reflected in the planets, suggesting mantles dominated by (Mg,Fe)O ferropericlase, and which have little water storage capacity.
The measurement of Th in Barnard's Star unlocks an analysis of the radiogenic heating in the planets: a combination of radioactive decay and GCE modelling implies that the specific radiogenic heat produced in their mantles is less than half that of Earth.
Scaling relations imply that all four planets are tidally locked, and the mantle mineralogy and low radiogenic heating suggest cool mantle temperatures and little capacity for volcanism and outgassing.
The contribution of tidal heating due to a potential 4:3 mean-motion resonance chain is however uncertain.
The planets' old age, proximity to the star, and modelled geothermal state imply minimal prospects for primary or secondary atmospheres, though next-generation facilities (in particular ANDES/ELT) will be able to test this.

Much of the investigation in this work can be applied more broadly than this individual system.
The interior modelling is based almost entirely on a stellar composition, which is usually accessible from stellar spectroscopy.
For the case of RV-discovered planets, a large number of high-resolution spectra are measured by necessity; but the stellar abundances are not routinely extracted, even though these contain important information on the nature of the planets.
With PLATO, \textit{Gaia} DR5, Earth~2.0, and the Nancy Grace Roman Space Telescope promising a surge in exoplanet detections in the next few years, stellar spectroscopy will be a vital tool in helping to constrain what these planets are made of.

\section*{Acknowledgements}

We thank the two anonymous reviewers, who both made numerous important insights to greatly improve several aspects of this manuscript.
We also acknowledge productive conversations with Yuqi Li, Richard Anslow, Adam Rains, Claudia Aguilera-G\'omez, Paula Jofr\'e and Sarah G.\ Kane.

This work has made use of data from the European Space Agency (ESA) mission {\it Gaia} (\url{https://www.cosmos.esa.int/gaia}), processed by the {\it Gaia} Data Processing and Analysis Consortium (DPAC, \url{https://www.cosmos.esa.int/web/gaia/dpac/consortium}).
Funding for the DPAC has been provided by national institutions, in particular the institutions participating in the {\it Gaia} Multilateral Agreement.

In addition to Python packages referenced in the text, we also acknowledge the use of \textsc{Astropy} \citep{astropy1, astropy2, astropy3}, \textsc{Matplotlib} \citep{matplotlib}, \textsc{NumPy} \citep{numpy}, and \textsc{pandas} \citep{pandas1, pandas2}.

CMG acknowledges the support of the UK Science and Technology Facilities Council (grant no. ST/W000903/1) and the ETH Postdoctoral Fellowship.
HSW acknowledges support from the Carlsberg Foundation through the FIRSTATMO project.

The authors declare no conflicts of interest.

\section*{Data Availability}

The data used in this work are all publicly available.
Analysis scripts will be made freely available upon acceptance of the manuscript at \url{https://github.com/xbyrne/barnard}.



\bibliographystyle{mnras}
\bibliography{bibliography}




\appendix

\section{A full \texorpdfstring{$N$}{}-body simulation of the Barnard's Star system}
\label{app:Nbody}

To assure ourselves of the stability predictions from \texttt{SPOCK}, we conduct one full $N$-body simulation of the Barnard's Star system, using \texttt{REBOUND} \citep{rebound} and the \texttt{WHFAST} integrator \citep{whfast}.
In the simulation, the planets are given masses $2.5\times$ their minimum masses, in accordance with the stability constraints from Section~\ref{sec:stability}.
The planets are set on circular orbits, starting from example mean longitudes of $(\lambda_{\rm d}, \lambda_{\rm b}, \lambda_{\rm c}, \lambda_{\rm e}) = (0, \pi/2, \pi, 3\pi/2)$.
The system is simulated for the same period of time over which \texttt{SPOCK} estimates the dynamical stability -- $10^9\times P_{\rm d}\approx 6.4~\rm{Myr}$ -- with timesteps equal to $0.05 P_{\rm d}$.

\begin{figure*}
\includegraphics[width=\textwidth]{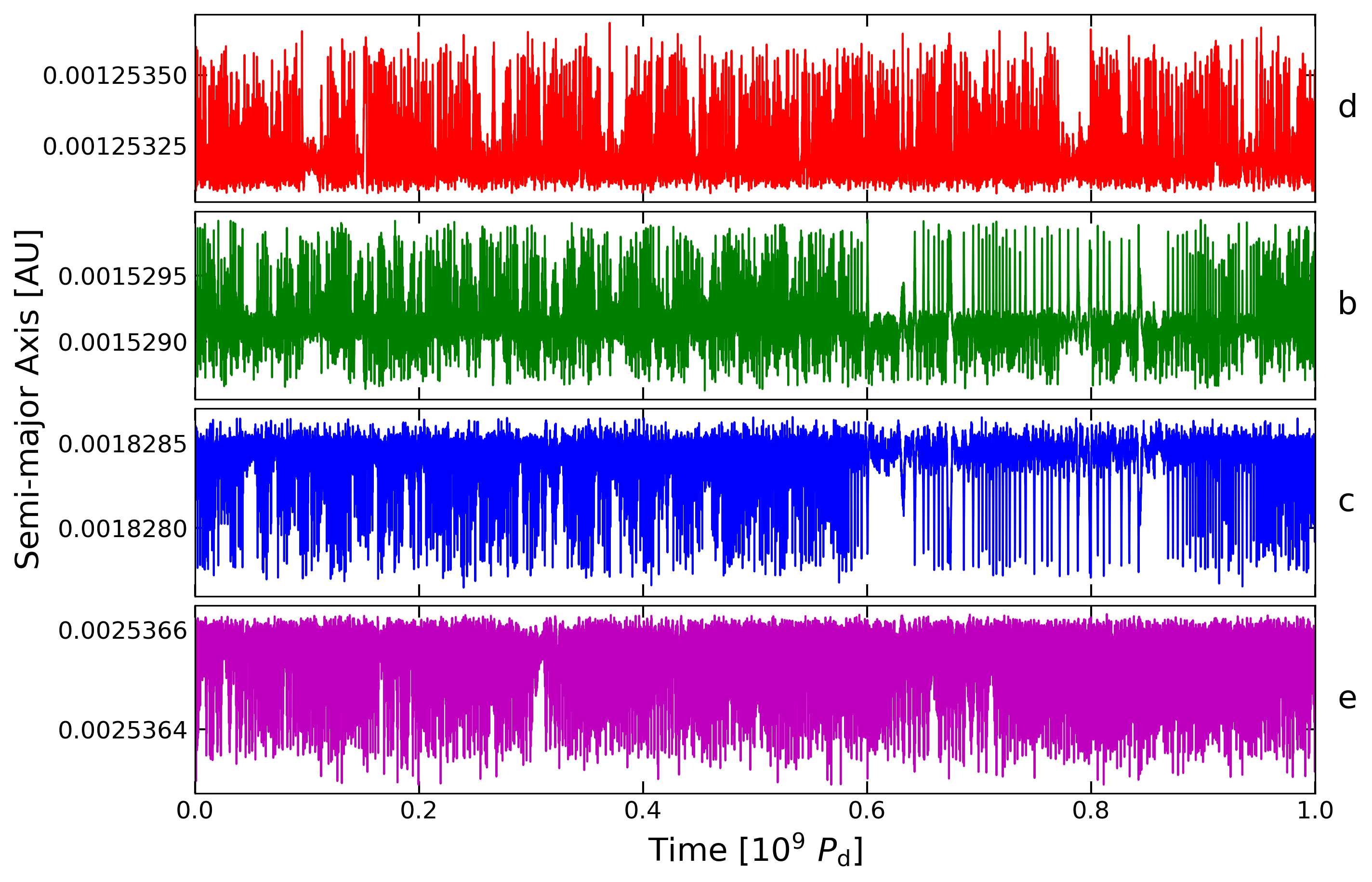}
\caption{
Evolution of semi-major axes during an example $N$-body simulation of the Barnard's Star system, in which each planet has a mass $2.5\times$ its minimum mass.
The panels are in order of orbital distance (d, b, c, e).
The planets' semi-major axes vary only on the order of $10^{-5}~{\rm au}$, without any instabilities.
}
\label{fig:Nbody}
\end{figure*}

The evolution of the planets' semi-major axes is shown in Figure~\ref{fig:Nbody}.
We see that, although the planets' orbital radii oscillate slightly, these changes are on the order of $10^{-5}~{\rm au}$ and the system is clearly stable, at least for the time period investigated.

\section{Abundance approximations for condensation modelling}
\label{app:ggchem_abunds}

The condensation of a Barnard's-Star-composition gas is simulated in Section~\ref{sec:ggchem}.
Not all elements are measured in Barnard's Star, so the abundances of some elements need to be approximated.
These approximations follow \citet{spaargaren25}, and are listed below.

He, a noble gas, does not directly impact chemistry; we use the solar abundance from \citet{asplund21}.

S has similar nucleosynthetic pathways to Si \citep{pignatari16, kobayashi20}, and shows very similar stellar abundances \citep{chen02}.
We thus assume the S abundance of Barnard's Star to equal that of Si.

For N, whose nucleosynthesis is simultaneous with that of O, we use the relation in \citet[][their equation~3]{nicholls17}, using the O abundance found in \citetalias{jahandar24}.

Ni has been found to be less abundant than Fe by an approximately constant ratio of $17.7\pm2.0$ in FGK stars \citep{wang19}; we find this to approximately hold among M dwarfs in the Hypatia catalogue \citep{hinkel14}.
We therefore use $\mathrm{Fe/Ni}=17.7$ to estimate the Ni abundance.


\bsp	
\label{lastpage}
\end{document}